\documentclass[aps,showpacs,showkeys,preprint]{revtex4}
\usepackage{epsfig,amssymb,bm,graphics,color}
\usepackage{epstopdf}
\newcommand*{\fs}[1]{#1\!\!\!/}
\newcommand*{\fsk}{k\!\!\!/}
\newcommand*{\ee}{e^+e^-}

\begin{document} {\normalsize }

\title{Breit-Wheeler process in very short electromagnetic pulses}


\author{
A.I.~Titov$^{a,b,c}$,   B.~K\"ampfer$^{a,d}$, H.~Takabe$^{c}$, and
A.~Hosaka$^e$}
 \affiliation{
 $^a$Helmholtz-Zentrum  Dresden-Rossendorf, 01314 Dresden, Germany\\
 $^b$Bogoliubov Laboratory of Theoretical Physics, JINR, Dubna 141980, Russia\\
 $^c$Institute of Laser Engineering, Yamada-oka, Suita, Osaka 565-0871, Japan\\
  $^d$Institut f\"ur Theoretische Physik, TU~Dresden, 01062 Dresden, Germany\\
 $^e$Research Center of Nuclear Physics, 10-1 Mihogaoka Ibaraki,
 567-0047 Osaka, Japan}

\begin{abstract}
 The generalized Breit-Wheeler process, i.e.~the
 emission of $\ee$ pairs off a probe photon propagating through
 a polarized
 short-pulsed electromagnetic (e.g.\ laser) wave field,
 is analyzed.
 We show that the production probability is determined
 by the interplay of two dynamical effects. The first one is
 related to the shape and duration of the pulse and the
 second one is the non-linear dynamics of the
 interaction of $e^\pm$ with the strong
 electromagnetic field.
 The first effect
 manifests itself most clearly in the weak-field regime,
 where the small field intensity is compensated by the
 rapid variation of the electromagnetic field in a limited
 space-time region, which intensifies the few-photon events and
 can enhance the production  probability by orders
 of magnitude compared to
 an infinitely long pulse.
 Therefore, short pulses may be considered as a powerful
 amplifier.
 The non-linear dynamics in the multi-photon Breit-Wheeler regime
 plays a decisive role
 at large field intensities,
 where effects of the pulse shape and duration are less important.
 In the transition regime, both effects must be taken
 into account simultaneously.
 We provide suitable  expressions for the $\ee$ production probability
 for kinematic regions  which can be used in transport codes.
\end{abstract}


\maketitle

\section{introduction}

The rapidly progressing laser technology developments~\cite{Tajima}
offer opportunities
for investigations of quantum systems with short
and/or intense pulses~\cite{Piazza}.
Several fundamental processes of electron-photon
interactions in the nonlinear regime thus become accessible. Once these are
understood experimentally and theoretically, e.g.
within the framework of the standard  model
of particle physics or plain quantum electrodynamics (QED),
one can search for new
phenomena hinting also to "new physics".
Among the elementary electromagnetic (e.m.)
interaction processes is the "conversion of light to matter".
Generically, this notion refers to the emergence of particles
coupling to the e.m.~field. Having in
mind electrons ($e^-$) and positrons ($e^+$)
one is interested in the conversion
rate into $e^\pm$, their phase space distributions,
the back-reaction on the original e.m.~field etc.

Several variants of such conversion processes are known. The linear
Breit-Wheeler
process~\cite{Breit-Wheeler-1934}
$\gamma' + \gamma \to e^+ + e^-$ refers to a perturbative QED
process; the generalization to the multi-photon process
$\gamma' + n \gamma \to e^+ + e^-$
(nonlinear Breit-Wheeler process) were done
in the pioneering work of
Reiss~\cite{Reiss} as well as Narozhny, Nikishov and Ritus
\cite{NR-64,Ritus-79}.
Attributing theses processes to
colliding null fields one can imagine another aspect. In the anti-node of
suitably counter propagating e.m.\ waves an
oscillating purely electric field can give rise
to the dynamical Schwinger effect~\cite{Blaschke-2013};
in the low-frequency limit one recovers
the famous Schwinger effect~\cite{Schwinger}
awaiting still its experimental
verification.
These kinds of pair creation processes
are related to highly non-perturbative effects~\cite{Dunne-2009,Hebenstreit-2011}.
Once pair production is
seeded in very intense fields further avalanche like particle
production can set in
which then could screen the original field or even limit
the attainable field strength~\cite{Fedotov-2010}.
One can relate the Breit-Wheeler process to the absorptive
part of the probe-photon correlator in an external e.m. field;
in our case the latter being a null field too.

In the present paper we focus on colliding null fields in the multi-photon
regime and consider the  generalized  Breit-Wheeler effect
for short pulses of e.m.\ wave fields
ranging from weak to high intensities.
Phrased differently we analyze $\ee$
pair production by a probe photon $\gamma'$ traversing a coherent e.m.\ (i.e.\ laser)
field. The latter one is characterized by the
reduced strength
\begin{eqnarray}
\xi^2=-\frac{e^2 \langle A^2  \rangle }{M^2_e}~,
\label{I-xi2}
\end{eqnarray}
where $\langle A^2 \rangle$ is the
mean square of the e.m.\ potential, and  $M_e$ is
the electron mass (we use natural units with
$c=\hbar=1$, $e^2/4\pi = \alpha \approx 1/137.036$).
A second relevant dimensionless variable characterizing both null fields is
\begin{eqnarray}
\zeta=\frac{s_{\rm thr}}{s},
\label{I-zeta}
\end{eqnarray}
where $s_{\rm thr}=4M_e^2$ and $s = 2 \omega \omega' (1 - \cos
\Theta_{\vec k \vec k'})$ (for head-on collision geometry,
$\Theta_{\vec k \vec k'}=\pi$); $\omega, \omega'$ and $\vec k,
\vec k'$ are the frequencies and three-wave vectors of the laser
field and the probe photon, respectively. The variable $\zeta$ is
a pure kinematic quantity with the meaning that for $\zeta > 1$
the linear Breit-Wheeler process $\gamma' + \gamma \to e^+ + e^-$
is sub-threshold, i.e.\ kinematically forbidden. However,
multi-photon effects enable the non-linear process $\gamma' + n
\gamma \to e^+ + e^-$ even for $\zeta > 1$ which we refer to as
sub-threshold pair production. The non-linear Breit-Wheeler
process has been experimentally verified in the experiment E-144
at SLAC~\cite{SLAC-1997}. There, the minimum number of photons
involved in one $\ee$ event
can be estimated by the integer part of $\zeta(1+\xi^2)$, i.e., five.
(To arrive at such an estimate recall that the reduced strength $\xi$ is related to
the laser intensity $I_L$ 
via $\xi^2 \simeq 0.56 (\omega(\rm eV))^{-2} 10^{-18} I_L /({\rm
W/cm}^2)$, and therefore, at  $\omega'=29$~GeV, $\omega=2.35$~eV,
and at peak focused laser intensity of $1.3 \times 10^{18} \, {\rm
W/cm}^2$, one gets $\xi =0.36$ and $\zeta=3.83$. The laser pulses
contained about thousand cycles in a shot, allowing to neglect the
details of the pulse shape and duration.)
  A laser intensity of $\sim 2\times 10^{22}$  W/cm${}^2$ has been already
  achieved~\cite{I-22}. Intensities of the order of
 $I_L \sim 10^{23}...10^{25}$ W/cm$^2$ are envisaged in near future  at
the  CLF~\cite{CLF}, ELI~\cite{ELI}, and HiPER~\cite{hiper} laser
facilities. Such large laser intensities allow for larger values
of $\xi^2 \sim I_L$ compared to the SLAC E-144 experiment.

The new generations of optical laser beams are expected to be essentially
 realized in short pulses (with femtoseconds duration) with only
 a few oscillations of the e.m.\ field.
High laser intensities are presently achieved by the chirped pulse amplification
 resulting in short pulses. As shown for the Compton effect
 in~\cite{Boca-2009,Heinzl-2009,Mackenroth-2011,Seipt-2011,Dinu,Seipt-2012}
 and for the Breit-Wheeler effect
in~\cite{TTKH-2012,Nousch,Krajewska,VillalbaChavez}
the pulse shape and the pulse duration become important.
That means the treatment of
the intense laser field as an infinitely long wave train
is no longer adequate. Keeping
the spatial plane-wave character we are going to explore here
the $\ee$ production as generalized
Breit-Wheeler
process in finite pulse approximation (FPA), i.e.\ investigate the impact of
the temporal pulse structure, and provide the conditions under which the
infinitely long pulse approximation (IPA) can be applied.
This problem is of practical interest for the investigation of
$\ee$ production in transport Monte Carlo
calculations~\cite{Elkina-2011,Bulanov-2011},
where the probability of pair production in a background field
is taken as an input.

We show below that the $\ee$ production probability is determined
by the non-trivial interplay of two dynamical effects. The first one is
related to the shape and duration of the pulse.
The second one is the non-linear dynamics of the
$e^\pm$ in the
strong
electromagnetic field, independently of the pulse
geometry.
These two effects play quite different roles in two limiting cases:
The pulse shape effects manifest most clearly
in the weak-field regime characterized by
small values of the product $\xi \,\zeta$.
The rapid variation
of the e.m. field in very short (sub cycle) pulses
enhances strongly few-photon events such that their
probability
may exceed the IPA prediction by
orders of magnitude.
Non-linear multi-photon dynamics of the strong
electromagnetic field plays
a dominant role at large values of $\xi^2$.
In the transition region, i.e.\
at intermediate values $\xi^2 \sim 1$, the pair
creation probability is determined
by the interplay of both effects which must be taken
into account simultaneously.

Our paper is organized as follows.
In Sect.~II we derive  the basic expressions  for
the probability of $\ee$ creation in FPA and consider a few prototypical
pulse envelope shapes.
In Sect.~III we discuss the case of ultra-short (sub cycle) pulses where the
number of
oscillations of the laser field is smaller than one.
The case of short pulses with a few oscillations
of the laser field within one pulse is considered in Sect.~IV.
In particular, we analyze the enhancement of the
production probability in the sub-threshold region at small values of $\xi^2$,
discuss the case of intermediate $\xi^2 \sim 1$, and evaluate
the production probability at large values of $\xi^2$.
Our conclusions are given in Sect.~V.
In Appendix~A, for completeness and
easy reference, we present some details of the derivation of
the production probability for very high intensities, $\xi^2\gg1$.

\section{electron-positron emission in a short pulse}

\subsection{General formalism}

 In the following we employ the e.m.~four-potential
 of the circularly polarized
 laser field in the axial gauge $A^\mu=(0,\,\vec A(\phi))$
 with
  \begin{eqnarray}
\vec{A}(\phi)=f(\phi) \left( \vec a_x\cos(\phi+\hat\phi)+
\vec a_y\sin(\phi +\hat\phi)\right)~,
\label{III1}
\end{eqnarray}
where $\phi=k\cdot x$ is invariant phase with four-wave vector
$k=(\omega, \vec k)$, obeying the null field property $k^2=k\cdot
k=0$ (a dot between four-vectors indicates the Lorentz scalar
product); $\hat\phi$ is the carrier envelope phase; $|\vec
a_x|^2=|\vec a_y|^2 = a^2$, $\vec a_x \vec a_y=0$. Transversality
means $\vec k \vec a_{x,y}=0$ in the present gauge. Instead
switching on/off the periodic e.m.~field we encode the finiteness
of a pulse in the envelope function $f(\phi)$ with
$\lim\limits_{\phi\to\pm\infty}f(\phi)=0$ (FPA). To characterize
the pulse duration one may use the number $N$ of cycles in a
pulse, $N=\Delta/\pi=\frac12\tau\omega$, where the dimensionless
quantity $\Delta$ or the duration of the pulse $\tau$ are further
useful measures. Below we analyze the dependence of observables on
the shape of $f(\phi)$ for a variety of relevant envelopes. The
IPA case is defined by $f(\phi)=1$. The carrier envelope phase
$\hat\phi$ is particularly important if it is comparable with the
pulse duration $\Delta$. In IPA it is anyhow irrelevant; in FPA
with $\hat\phi\simeq\Delta$ the production probability would be
determined by an involved interplay of the carrier phase, the
pulse duration and pulse shape as well as the parameters $\xi$ and
$\zeta$ as emphasized, e.g., in
\cite{Krajewska,Hebenstreit,Mackenroth}.
In present work, we drop the carrier phase, thus assuming
$\hat\phi\ll\Delta$, and concentrate on the dependence of the
production probability on the parameters $\xi$ and $\zeta$
together with pulse shape and pulse duration. A detailed analysis
of the impact of $\hat\phi$ on the pair production needs a
separate investigation which is postponed to subsequent work.

Utilization of the e.m.~potential of (\ref{III1}) leads to two significant modifications
of the transition amplitude in FPA compared to IPA.
In IPA, the Volkov solutions~\cite{Volkov,LL-1982} refer to Fermions with
quasi-momenta and dressed masses.
In FPA, all in- and out-~momenta and masses
take their vacuum values.
The finite (in space-time) e.m. potential (\ref{III1}) for FPA
requires the use of Fourier integrals
for invariant amplitudes, instead of Fourier
series which are employed in IPA.
The partial harmonics
become thus continuously in FPA.
The $S$ matrix element is expressed generically as
\begin{eqnarray}
S_{fi}
=\frac{-ie}{\sqrt{2p_02p_0'2\omega'}}
\int\limits_\zeta^\infty dl
\, M_{fi}(l)(2\pi)^4\delta^4(k'+ lk -p-p'),
\label{III3}
\end{eqnarray}
where $k$, $k'$,  $p$ and $p'$ refer to the four-momenta of the
background (laser) field (\ref{III1}),
incoming probe photon, outgoing positron and electron,
respectively.
The transition matrix $ M_{fi}(l)$, similarly to the case
of the non-linear Compton effect
\cite{Boca-2009,Mackenroth-2011,Seipt-2011,Dinu},
consists of four terms
\begin{eqnarray}
\, M_{fi}(l)=\sum\limits_{i=0}^3  M^{(i)}\,C^{(i)}(l)~,
\label{III4}
\end{eqnarray}
where
\begin{eqnarray}
C^{(0)}(l)&=&\frac{1}{2\pi}\int\limits_{-\infty}^{\infty}
d\phi \,{\rm e}^{il\phi -i{\cal P(\phi)}}~,\nonumber\\
C^{(1)}(l)&=&\frac{1}{2\pi}\int\limits_{-\infty}^{\infty}
d\phi f^2(\phi)\,{\rm e}^{il\phi -i{\cal P(\phi)}}~,\nonumber\\
C^{(2)}(l)&=&\frac{1}{2\pi}\int\limits_{-\infty}^{\infty}
d\phi f(\phi)\,\cos\phi\,{\rm e}^{il\phi -i{\cal P(\phi)}}~,\nonumber\\
C^{(3)}(l)&=&\frac{1}{2\pi}\int\limits_{-\infty}^{\infty}
d\phi f(\phi)\,\sin\phi\,{\rm e}^{il\phi -i{\cal P(\phi)}}~,
\label{III5}
\end{eqnarray}
with
\begin{eqnarray}
{\cal P(\phi)}=z\int\limits_{-\infty}^{\phi}\,d\phi'\,\cos(\phi'-\phi_0)f(\phi')
-\xi^2\zeta u\int\limits_{-\infty}^\phi\,d\phi'\,f^2(\phi')~.
\label{III6}
\end{eqnarray}
The quantity $z$ is related to $\xi$, $l$,
$u\equiv(k'\cdot k)^2/\left(4(k\cdot p)(k\cdot p')\right)$, and $u_l\equiv l/\zeta$ via
\begin{eqnarray}
z=2l\xi\sqrt{\frac{u}{u_l}\left(1-\frac{u}{u_l}\right)}~.
\label{III7}
\end{eqnarray}
The phase $\phi_0$ is equal to the azimuthal angle of
the electron emission direction in the $\ee$ pair rest frame $\phi_{p'}$ and
is related to the azimuthal angle of
the positron as $\phi_0=\phi_p + \pi$.
Similarly to IPA,
it can be determined through invariants $\alpha_{1,2}$
as $\cos\phi_0=\alpha_1/z$,  $\sin\phi_0=\alpha_2/z$ with
$\alpha_{1,2}=e\left(a_{1,2}\cdot p/k\cdot p-a_{1,2}\cdot p'/k\cdot p'\right)$.

The transition operators $ M^{(i)}$ in Eq.~(\ref{III4})
have the form
\begin{eqnarray}
M^{(i)}=\bar u_{p'}\,\hat M^{(i)}\,v_p~
\label{B1}
\end{eqnarray}
with
\begin{eqnarray}
 \hat M^{(0)}&=&\fs\varepsilon'~,\qquad
 \hat M^{(1)}=\frac{e^2({\fs a_1}\,\fsk\fs\varepsilon'\,\fsk\fs{ a_1}
 + \fs{a_2}\,\fsk\fs\varepsilon'\,\fsk\fs{a_2})}
 {4(k\cdot p)(k\cdot p')} ~,\nonumber\\
\hat M^{(2)}&=&\frac{e\fs a_1\fs k\fs \varepsilon'}{2(k\cdot p')}
+
\frac{e\fs \varepsilon'\fs k\fs a_1}{2(k\cdot p)}~,\qquad
\hat M^{(3)}=\frac{e\fs a_2\fs k\fs \varepsilon'}{2(k\cdot p)}
+
\frac{e\fs \varepsilon'\fs k\fs a_2}{2(k\cdot p')}~,
\label{B2}
\end{eqnarray}
where $u$ and $v$ are Dirac spinors of the electron and positron,
respectively, and
$\varepsilon'$ is the polarization four-vector of the probe photon.

The integrand of the function $C^{(0)}$ does not contain the envelope
function $f(\phi)$ and therefore
it is divergent. One can regularize it by using the
prescription of Ref.~\cite{Boca-2009}.  The formal result
\begin{eqnarray}
C^{(0)}(l)=\frac{1}{2\pi l} \int\limits_{-\infty}^{\infty}
d\phi
\left( z\cos(\phi-\phi_0)\,f(\phi)-\xi^2\zeta u\,f^2(\phi)\right)
\,{\rm e}^{il\phi -i{\cal P(\phi)}}
+\delta(l)\,{\rm e}^{-i{\cal P}(0)}
\label{III8}
\end{eqnarray}
contains a singular term (last term) which however does not contribute because of kinematical
considerations implying $l > 0$.
The differential probability of $\ee$ pair production
in terms of the transition matrix $ M_{fi}(l)$ in Eq.~(\ref{III3})
reads
\begin{eqnarray}
{d W}
=\frac{\alpha\zeta^{1/2}}{2\pi N_0 M_e}
 \,\int\limits_\zeta^{\infty}
 \,dl\,\, |M_{fi}(l)|^2\,\frac{d\vec p}{2p_0}\,\frac{d\vec p'}{2p'_{0}}
 \delta^4(k'+ lk -p-p')~.
\label{III9-0}
\end{eqnarray}
It may be represented in conventional form as
a function of $u$ and $\phi_p$
\begin{eqnarray}
\frac{d W}{d\phi_p\,du }
=\frac{\alpha M_e\zeta^{1/2}}{16\pi N_0}
\,\frac{1}{u^{3/2}\sqrt{u-1}}
 \,\int\limits_\zeta^{\infty}\,dl\ w{(l)}
\label{III9}
\end{eqnarray}
with
\begin{eqnarray}
\frac12\,w(l)&=&
(2u_l+1)|C^{(0)}(l)|^2 +\xi^2(2u-1)(|C^{(2)}(l)|^2 +|C^{(3)}(l)|^2 )
\nonumber\\
&+&
{\rm Re}\, C^{(0)}(l)\left(
\xi^2 {C^{(1)}(l)} -\frac{2z}{\zeta}(\alpha_1 {C^{(2)}(l)}
+\alpha_2  {C^{(3)}(l)})
\right)^*~.
\label{III20}
\end{eqnarray}
 The differential  probability $dW$ in Eq.~(\ref{III9}), in fact is the probability per
 unit time (or rate).
 The time units in IPA ($\Delta T^{(IPA)}$) and FPA ($\Delta T^{(FPA)}$) are different.
 The ratio of $N_0\equiv \Delta T^{(IPA)}/\Delta T^{(FPA)}$
 may be evaluated as following. The variation of e.m. energy of a
 pulse in a volume $\int_{-\infty}^{\infty}S\,dz$, where $S$
 is an unit cross section in the $ x - y$ plane, per $\Delta T^{(FPA)}$
 is equal to the integral of the energy flux vector for the
 electromagnetic energy
 $ \vec I$ (Poynting vector)
 over the area $\oint_s d\vec S\,\vec I$. Taking into account that
 this integral is the same in FPA and IPA
 one finds
 \begin{eqnarray}
 N_0 =\frac{\int\limits_{-\infty}^{\infty}S\,dz\,
 ({\mathbf{E}^2_{FPA}}+{\mathbf{B}^2_{FPA}})}
  {\int\limits_{0}^{\lambda}S\,dz\,
 ({\mathbf{E}^2_{IPA}}+{\mathbf{B}^2_{IPA}})}
 =\frac{1}{2\pi} \int \limits_{-\infty}^{\infty}
d\phi\, (f^2(\phi) +f^{'2}(\phi)),
 \label{T1}
 \end{eqnarray}
 where $\lambda=2\pi/\omega$ is the wave length, and
 $\mathbf{E}=-\frac{\partial \mathbf{A}}{dt}$ and
 $\mathbf{B}={\mathbf{\nabla}}\times \mathbf{A}$
 are electric and magnetic fields, respectively.
 For a convenient comparison
 of IPA and FPA results,
 the latter one is scaled in Eq.~(\ref{III9}) by $1/N_0$. For IPA, $N_0=1$.


 \subsection{Envelope functions}

 We consider one-parameter and two-parameter envelope functions.
 Among the one-parameter functions we choose
 the hyperbolic secant~(hs) and  Gaussian~(G)
 pulses~\cite{TTKH-2012,Seipt-2011}
\begin{eqnarray}
f_{\rm hs}(\phi)=\frac{1}{\cosh\frac{\phi}{\Delta}}~,\qquad
f_{\rm G}(\phi)=\exp\left[-\frac{\phi^2}{2\tau^2_{\rm G}}\right]~.
\label{E1}
\end{eqnarray}
As the two-parameter function, we choose the symmetrized
Fermi~(sF) shape~\cite{Luk}
\begin{eqnarray}
f_{\rm sF}(\phi)=\frac{\cosh\frac{\tau_{\rm sF}}{b} +1}
{\cosh\frac{\tau_{\rm sF}}{b} +\cosh{\frac{\phi}{b}}}~.
\label{E2}
\end{eqnarray}

The  scale parameters $\Delta$, $\tau_{\rm G}$ and  $\tau_{\rm sF}$
determine the normalization factor $N_0$ in (\ref{T1}):
$N_{\rm hs}=\frac{\Delta}{\pi}\left(1+  \frac{1}{3\Delta^2} \right)$,
$N_{\rm G}=\frac{\tau_{\rm G}}{2\sqrt{\pi}}\left(1 + \frac{1}{2\tau^2_{\rm G}}\right)$,
and
$N_{\rm sF}=\frac{b}{\pi}\left( F_1\left(t \right)
\log\frac{1+\exp[\tau_{\rm sF}/b]} {1+\exp[-\tau_{\rm sF}/b]}
 +  F_2\left(  t \right) \right)$
 with
 $t =({ \cosh\frac{\tau_{\rm sF}} {b} } +1 )/{\sinh\frac{\tau_{\rm sF}} {b}}$.
In the latter case we have defined
$F_1(t)=(t^2+1)(-t^4 +10 t^2 -1)/16t$ and
$F_2(t)=
\left(  3t^{10}-  35t^8+  90t^6-  90t^4+  35t^2-  3\right)/({24(t^2-1)^3})$.
In the limit  of small ${b}/{\tau_{\rm sF}}\to 0 $, one finds
$N_{\rm sF} = \frac{\tau_{\rm sF}}{\pi}
\left(1-\frac{5}{6}\frac{b}{\tau_{\rm sF}}  \right)
+ {\cal O}\left(\exp\left[-\frac{\tau_{\rm sF}}{b}\right] \right)$.

For large $\Delta$ and $\tau_{\rm sF}$, and small ${b}/{\tau_{\rm sF}}\ll 1$,
one can find $N_{\rm sF}\simeq N_{\rm hs}$ at $\tau_{\rm sF}=\Delta$.
Therefore, for the sake of comparison we denote $N_0=N_{\rm hs}$, $\tau_{sF}=\Delta$
and choose the ratio
${b}/\Delta$ as an independent parameter, which determines in turn
the normalization
factor  $N_{\rm sF}$ at finite values of ${b}/\Delta$.

The scale parameters $\Delta$ and $\tau_{\rm G}$
are related to each other
by $\tau_{\rm G}=\sqrt{\pi} N_0\left( 1+\sqrt {1-\frac{1}{2\pi N_0^2} } \right)$
for the fixed normalization factors $N_{0}=N_{G}=N_{\rm hs}$.

\begin{figure}[h!]
\includegraphics[width=0.35\columnwidth]{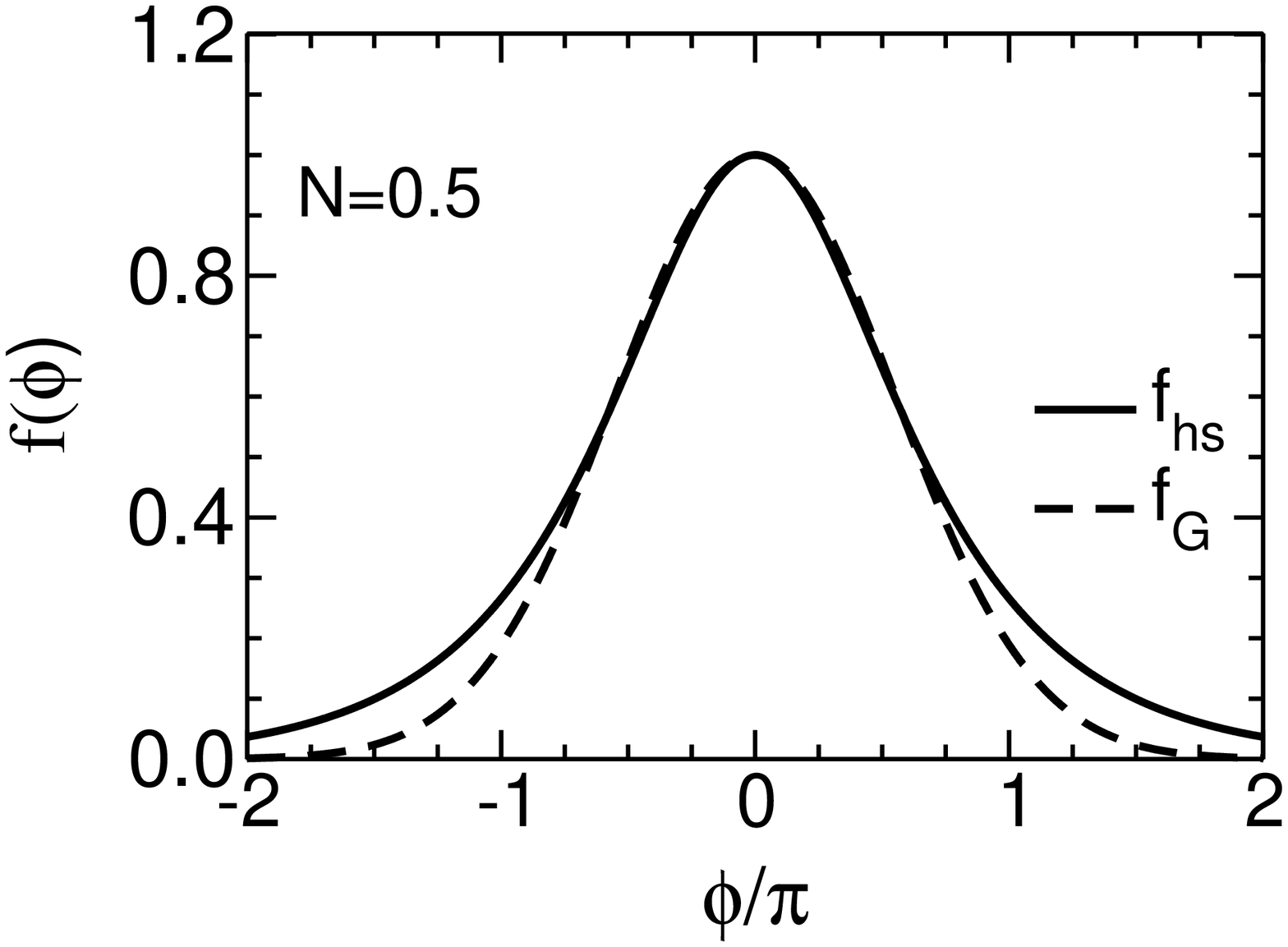}\qquad
\includegraphics[width=0.35\columnwidth]{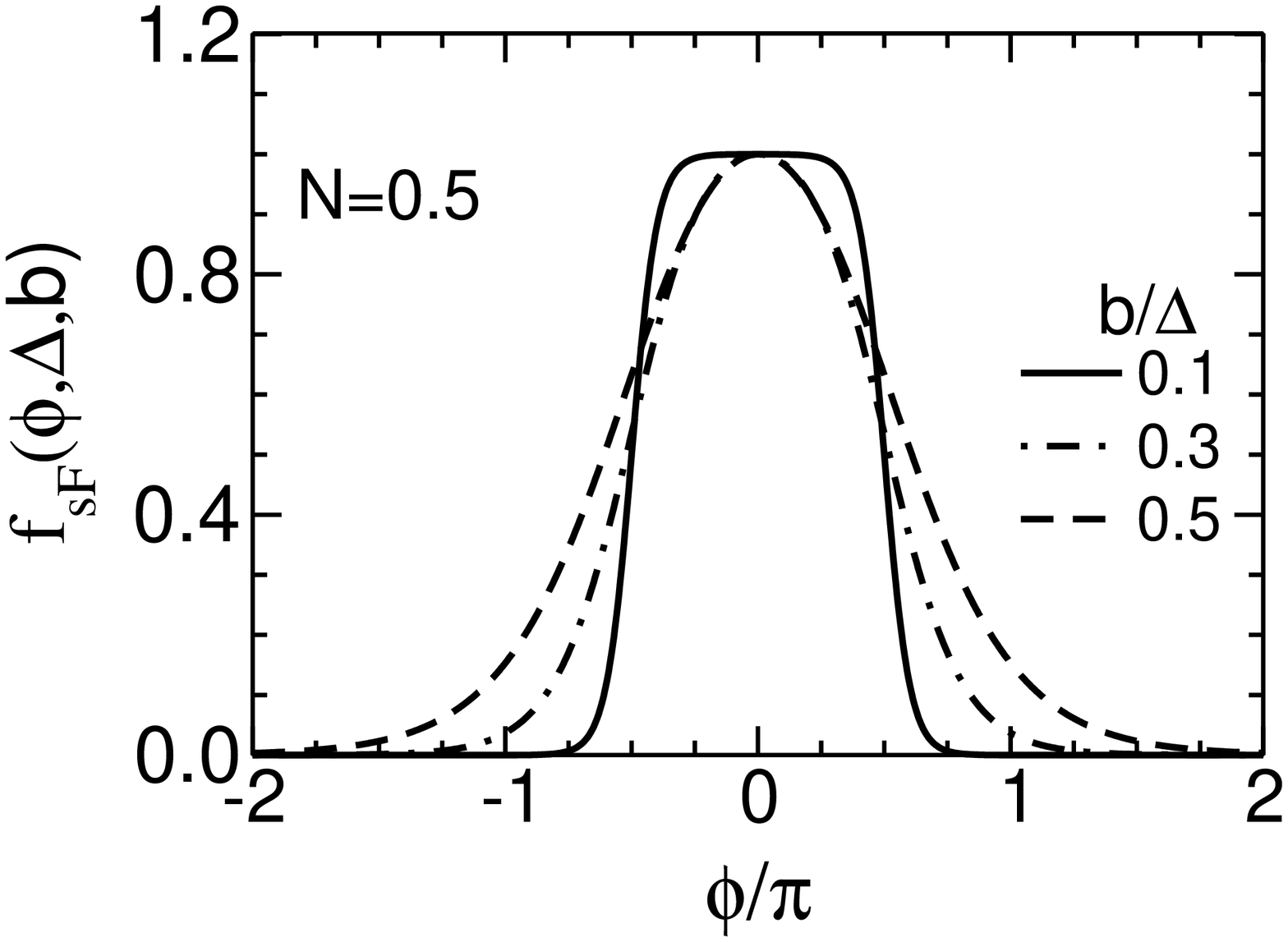}\\
\vspace*{1mm}
\includegraphics[width=0.35\columnwidth]{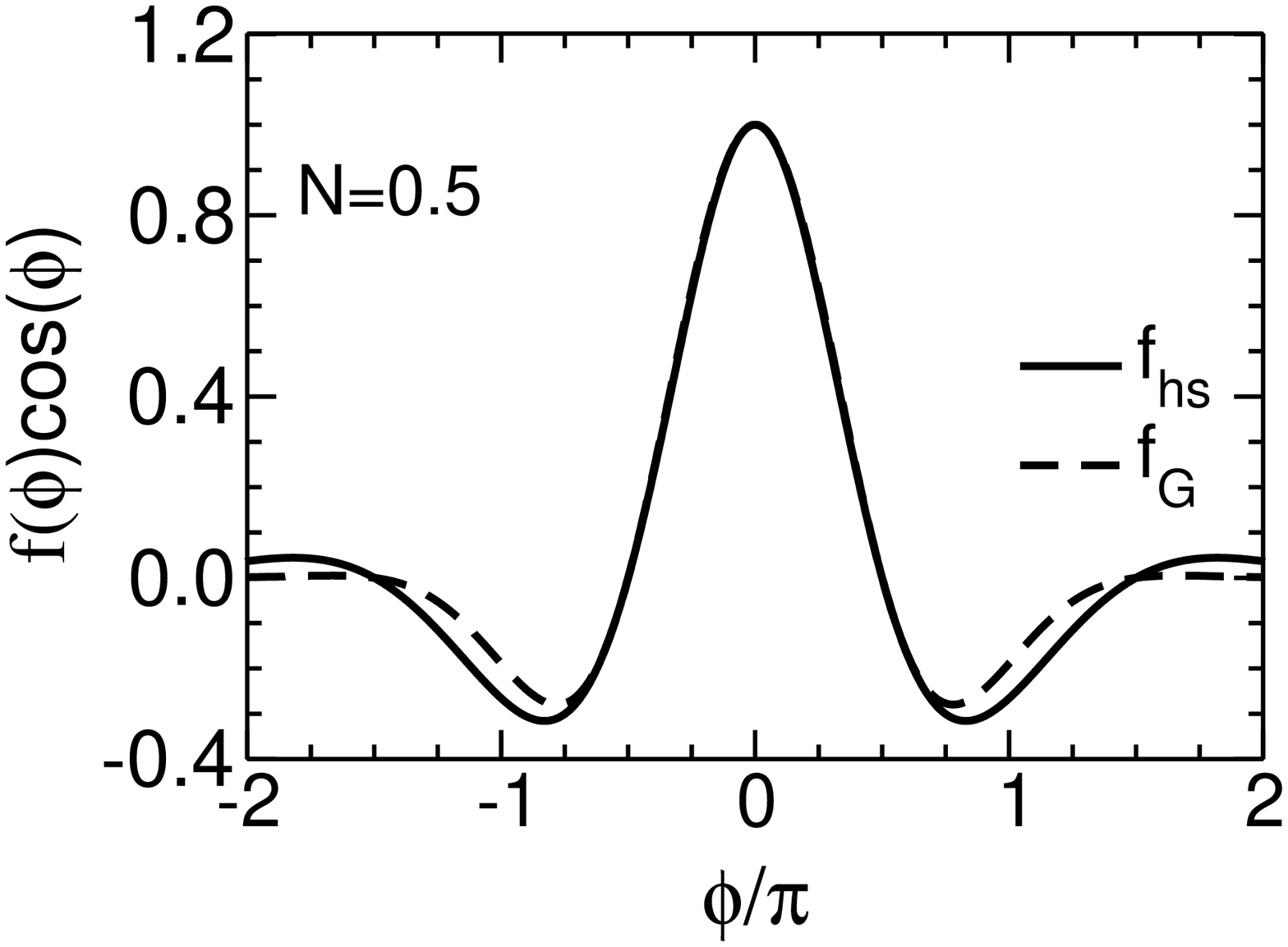}\qquad
\includegraphics[width=0.35\columnwidth]{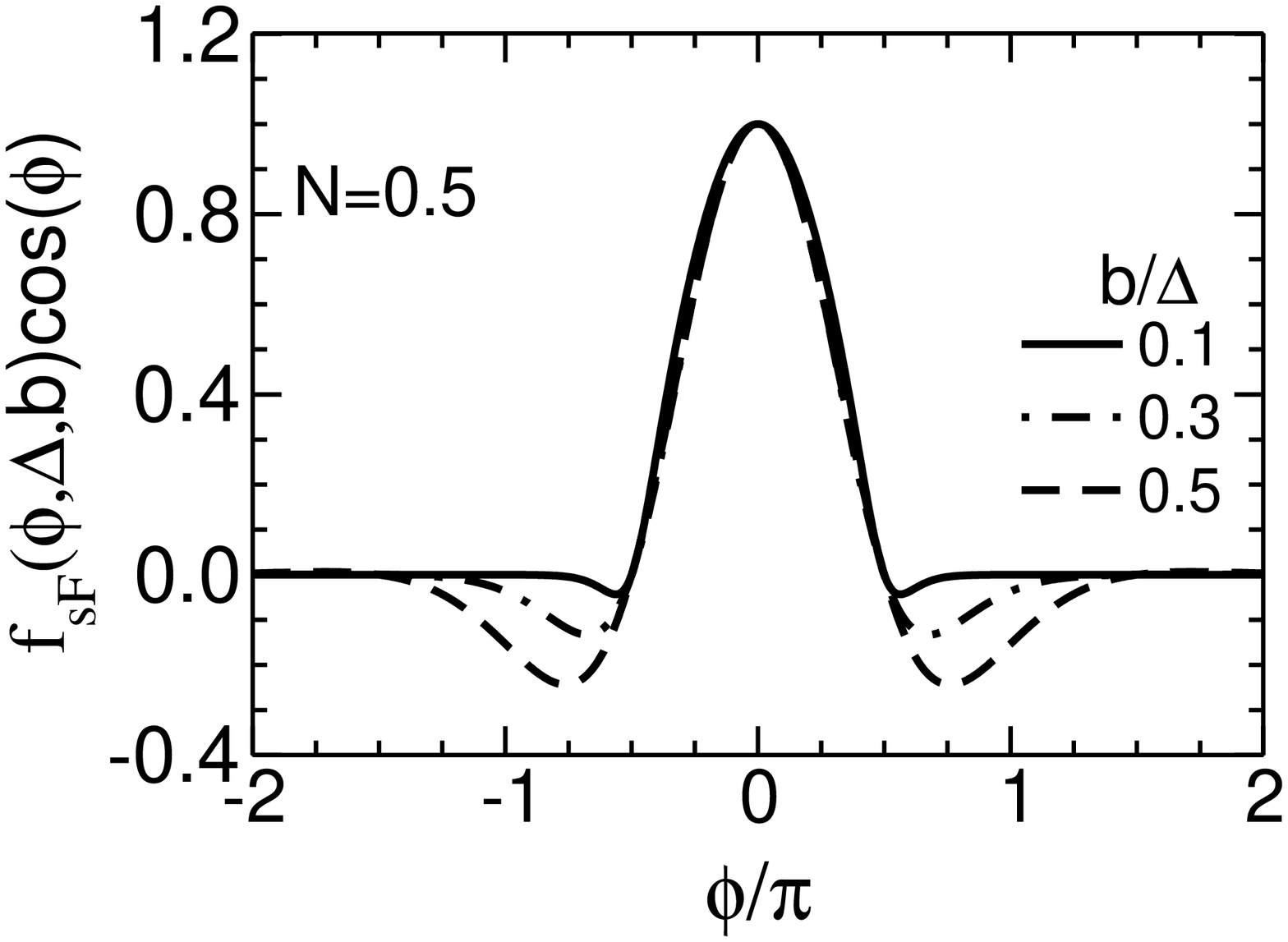}\\
\vspace*{1mm}
\includegraphics[width=0.35\columnwidth]{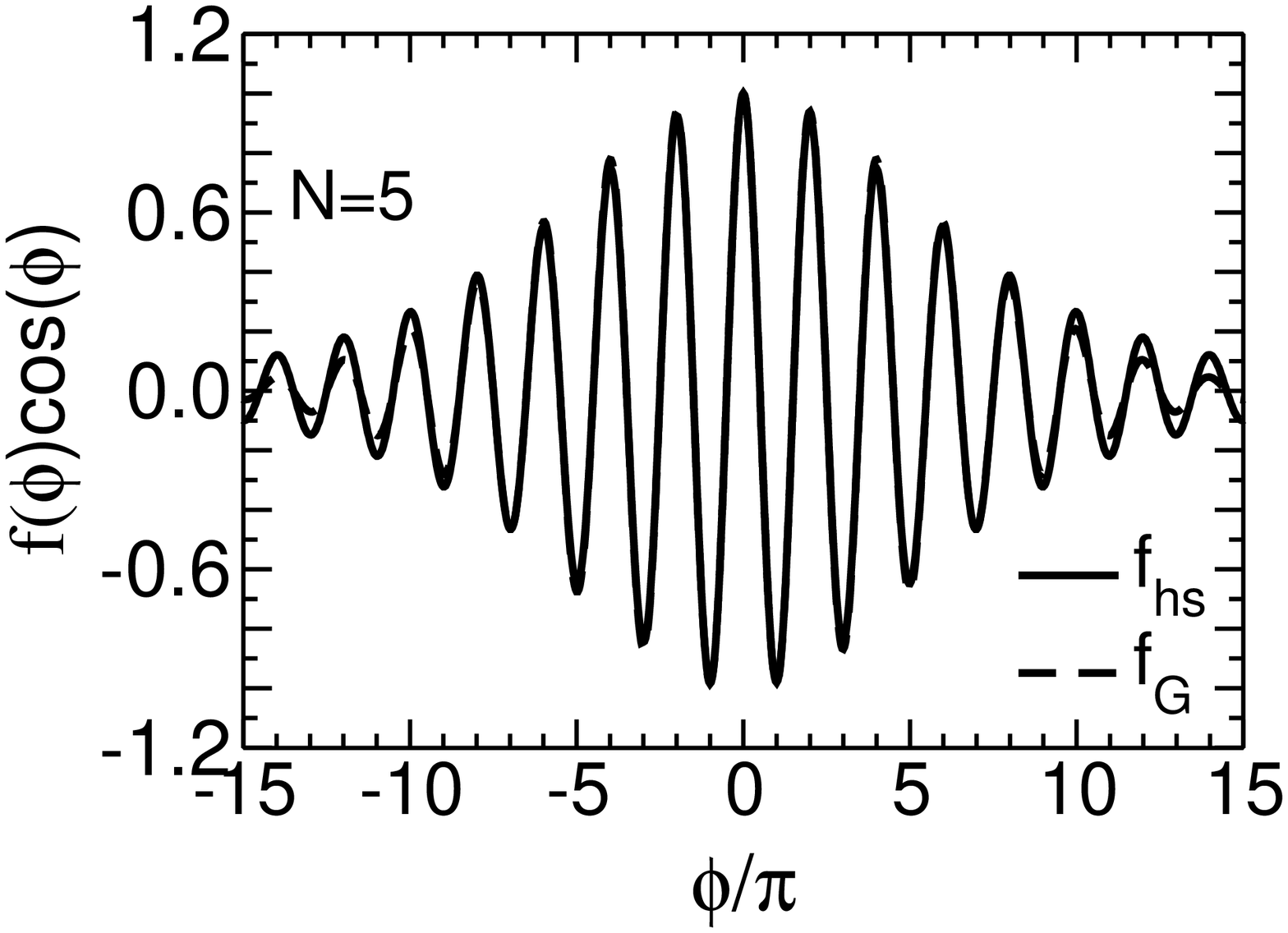}\qquad
\includegraphics[width=0.35\columnwidth]{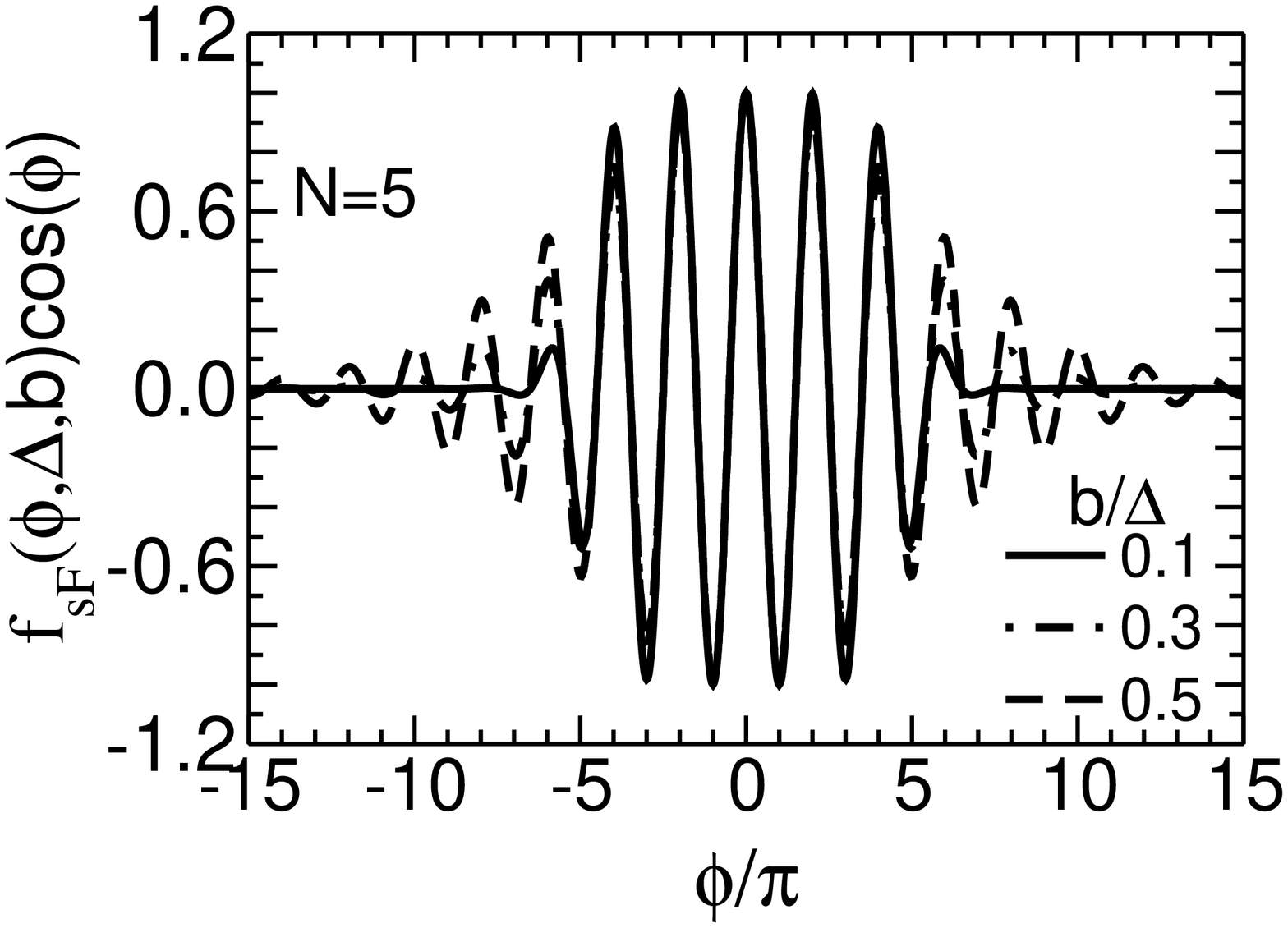}
\caption{\small{
Pulse envelope $f(\phi)$ (top panels) and the product $f(\phi)\cos\phi$
(middle and bottom  panels) as a function of the invariant phase $\phi$.
The left and right  panels correspond to the one- and two-parameter
envelope functions, respectively.
The top and the middle panels exhibit
an ultra-short pulse with the number of oscillations less than one, $N=0.5$,
while the bottom panels are for a short pulse with $N=5$.
 \label{Fig:1} }}
\end{figure}
The one- and two-parameter envelope functions are exhibited at the left and right
panels of  Fig.~\ref{Fig:1}.
Top panels depict to the envelope functions $f(\phi)$, the middle and bottom
panels depict the product $f(\phi)\cos\phi$, which determines
the function ${\cal P}$ in Eq.~(\ref{III6}).
The top and the middle panels are for
an ultra-short pulse (sub cycle) with the number of oscillations less than one, $N=0.5$.
The bottom panels correspond to a short pulse with $N=5$.
One-parameter envelopes are similar to each other and are close to the two-parameter envelope with
$b/\Delta\simeq 0.5$.  Decreasing $b/\Delta$ results in an essential
modification of the envelope function $f(\phi)$: it becomes close to the
flat-top profile with the double step
shape $\theta(\Delta^2 - \phi^2)$.

\section{Ultra-short pulses}

In this section we consider  the pair production due to
interaction of the probe photon
with an ultra-short pulse, where the number of cycles less than one.

\subsection{The case of small field intensity $(\xi^2\ll1)$ }

Consider first the case of small field intensities and a finite sub-threshold
parameter $\zeta$ characterized by the relations $z\ll 1$ or $\xi\zeta\ll1$.

The basic functions $C^{(i)}(l)$ in Eqs.~(\ref{III5}) and (\ref{III20}) can be expressed in this regime as a superposition of the functions
\begin{eqnarray}
{\cal Y}(l)=\frac{1}{2\pi}\int\limits_{-\infty}^{\infty}d\phi\,{\rm e}^{il\phi}\,f(\phi)\,g(\phi)
=\int\limits_{-\infty}^{\infty}dq\,F(l-q)\,G(q)~,
\label{U01}
\end{eqnarray}
where $F(p)$ and $G(q)$ are the Fourier transforms of the envelope function $f(\phi)$
and the  function $g(\phi)=\exp\left[ -i{\cal P}(\phi)\right]$, respectively:
$F(p)=\frac{1}{2\pi}\int\limits_{-\infty}^{\infty}
d\phi\,{\rm e}^{i p\phi}\,f(\phi)$,
$G(q)=\frac{1}{2\pi}\int\limits_{-\infty}^{\infty}d\phi\,
{\rm e}^{i q\phi}\,g(\phi)$.
For small values of $z$, $z\ll1$,  $G(q)\simeq\delta(q-q_0)$,
where $q_0\simeq \langle {\cal P'_{\phi}} \rangle$
with $q_0\sim \xi\zeta\ll 1$,  and  ${\cal Y}(l)\simeq F(l) $.
Keeping the leading terms in Eq.~(\ref{III20}) with
$C^{(i)}\simeq{\cal Y}(l-1)\simeq F(l-1)$, one can obtain an approximate
expression for the total production probability:
\begin{eqnarray}
W=
{\alpha M_e\zeta^{1/2}\xi^2}
\int\limits_\zeta^{\infty}dl\Phi(l)\,F^2(l-1)~,
\label{U2}
\end{eqnarray}
with
\begin{eqnarray}
\Phi(l)=v\int\limits_0^1 d\cos\theta\,\left(\frac{u}{u_l} -\frac{u^2}{u_l^2} + u -\frac12  \right)~,
\label{U3}
\end{eqnarray}
where $u=1/(1-v^2\cos^2\theta)$;  $\theta$ and $v$ are the polar angle and the velocity of the outgoing positron
in the $\ee$ c.m.s., respectively: $v=\sqrt{1-\zeta/l}$. An explicit calculation results in
\begin{eqnarray}
\Phi(l)=\frac12\left\{
\left( 1+\frac{\zeta }{l} - \frac{\zeta^2}{2l^2}\right)\log\frac{1+v}{1-v}
-v\left( 1+\frac{\zeta}{l}\right)
\right\}~.
\label{U4}
\end{eqnarray}

The Fourier transforms of the envelope functions (\ref{E1}) and (\ref{E2})
read
\begin{eqnarray}
F_{\rm hs}(l)&=&\frac{\Delta} {  2\cosh  {\frac12\pi\Delta  l}} ~,\nonumber\\
F_{\rm G}(l) &=&\frac{\tau_{\rm G}}{\sqrt{2\pi}}\exp\left[ -{\frac12\tau_{\rm G}^2l^2}  \right]~,\nonumber\\
F_{\rm sF}(l)&=&\frac{1+{\exp}\left[{-\frac{\Delta}{b}}\right] }
{1-{\exp}\left[{-\frac{\Delta}{b}}\right] }\,
\frac{ b\,\sin\Delta l } {\sinh \pi bl}~.
%
\label{U5}
\end{eqnarray}

\begin{figure}[h!]
\includegraphics[width=0.35\columnwidth]{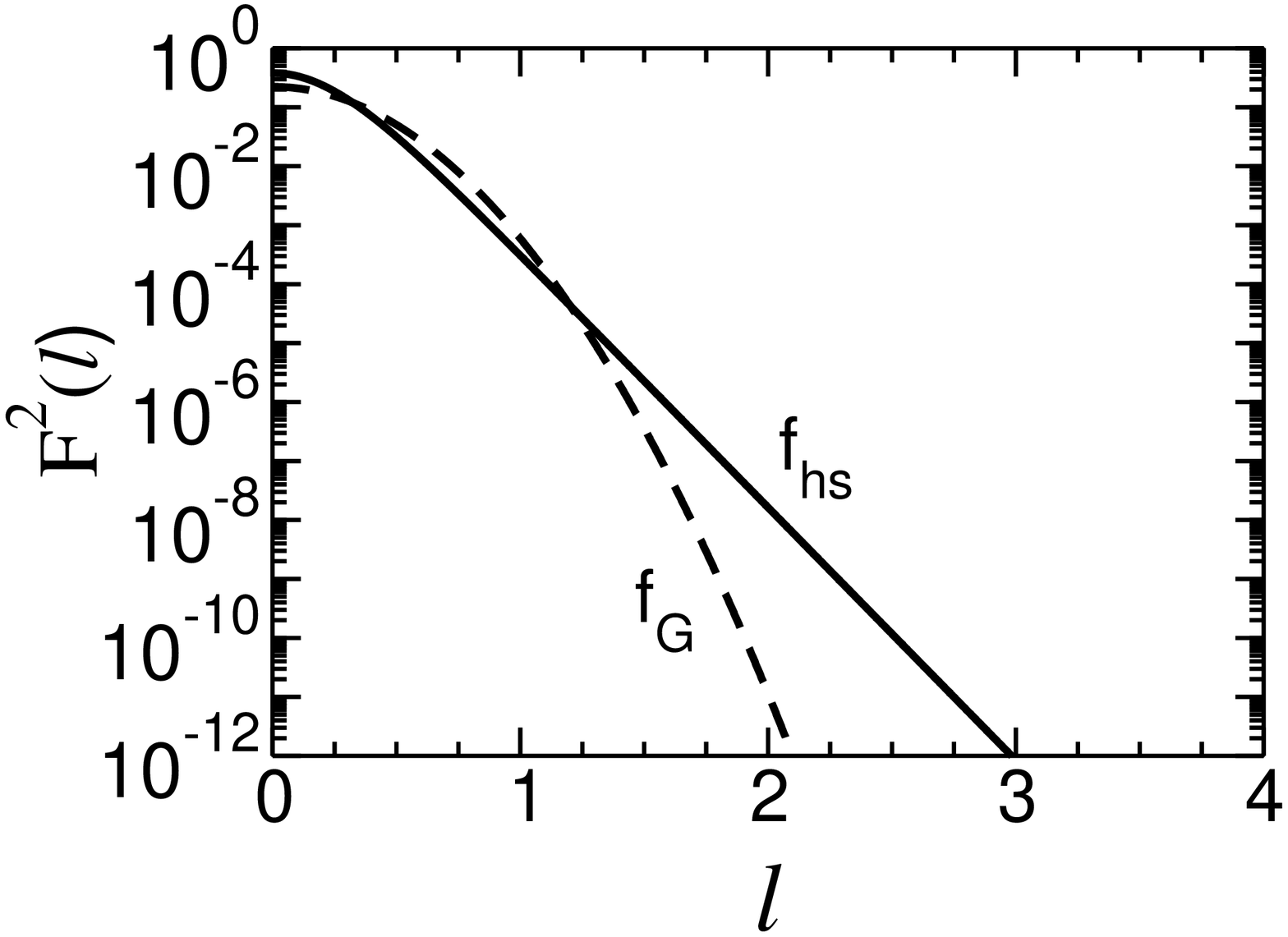}\qquad
\includegraphics[width=0.35\columnwidth]{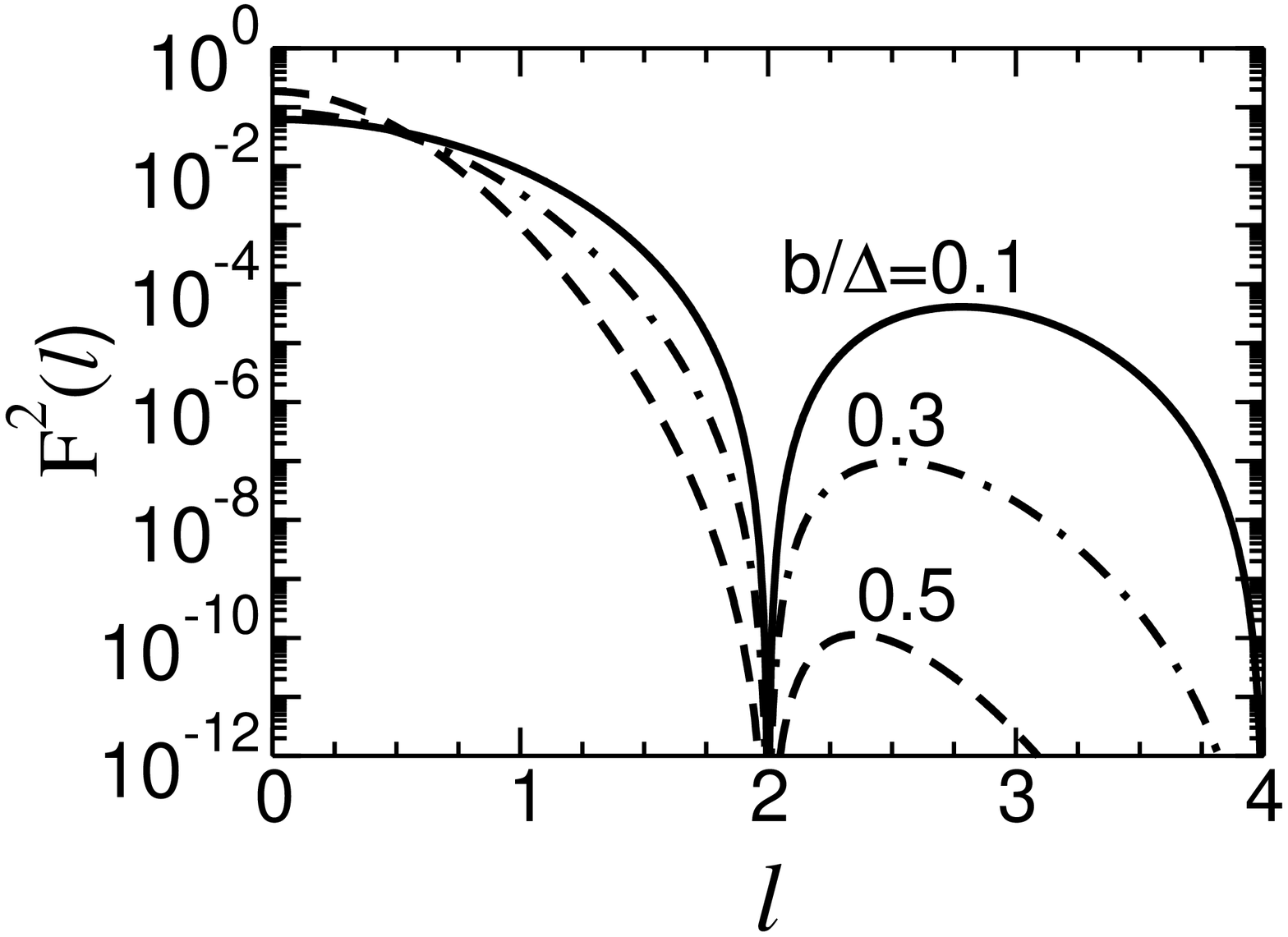}\\
\caption{\small{
Square of the Fourier transforms of envelope functions for a sub cycle pulse
with $N=0.5$. Left panel:
The solid and dashed curves depict the hyperbolic secant and Gaussian shapes,
respectively.
Right panel: The solid, dot-dashed and dashed curves show
the symmetrized Fermi shape for $b/\Delta=0.1$, 0.3, and 0.5,
respectively.
 \label{Fig:2} }}
\end{figure}

The square of the Fourier transforms of the envelope functions for a
sub cycle pulse with $N=0.5$ are presented in Fig.~\ref{Fig:2}.
On the left panel, the solid and dashed curves correspond to the hyperbolic secant
and Gaussian shapes, respectively. One can see a
fast monotonic decrease of both the functions
with some enhancement of $F_{\rm hs}$ at large values of  $l$.
The square of the Fourier transform for the symmetrized Fermi shape is shown in
the right panel, where the solid, dot-dashed and dashed curves correspond
to the ratios $b/\Delta=0.1$, 0.3, and 0.5,
respectively.
One can see large qualitative and quantitative
differences between the one-parameter and
symmetrized Fermi shapes, in particularly, at $b/\Delta\leq 0.3$.
In the second case, $F^2$ decreases exponentially as
$\exp\left[-2\pi\Delta\frac{b}{\Delta} \right]$. The slope decreases proportionally
to ${b}/{\Delta}$ (at fixed $\Delta$).
Also, the function oscillates with the period
$\delta\,l=\pi/\Delta=\pi/0.5\pi=2$.
Contrary to the above one-parameter shapes,
the function $F_{\rm sF}$ has a significant high-$l$ component at
$2\leq l\leq 4$. This strong effect is not seen in the $\phi$-space
(cf. Fig.~\ref{Fig:1}, top panels), where all envelope functions look
similar to each other. However, the differences in $l$-space
are very important for the pair production.

\begin{figure}[h!]
\includegraphics[width=0.35\columnwidth]{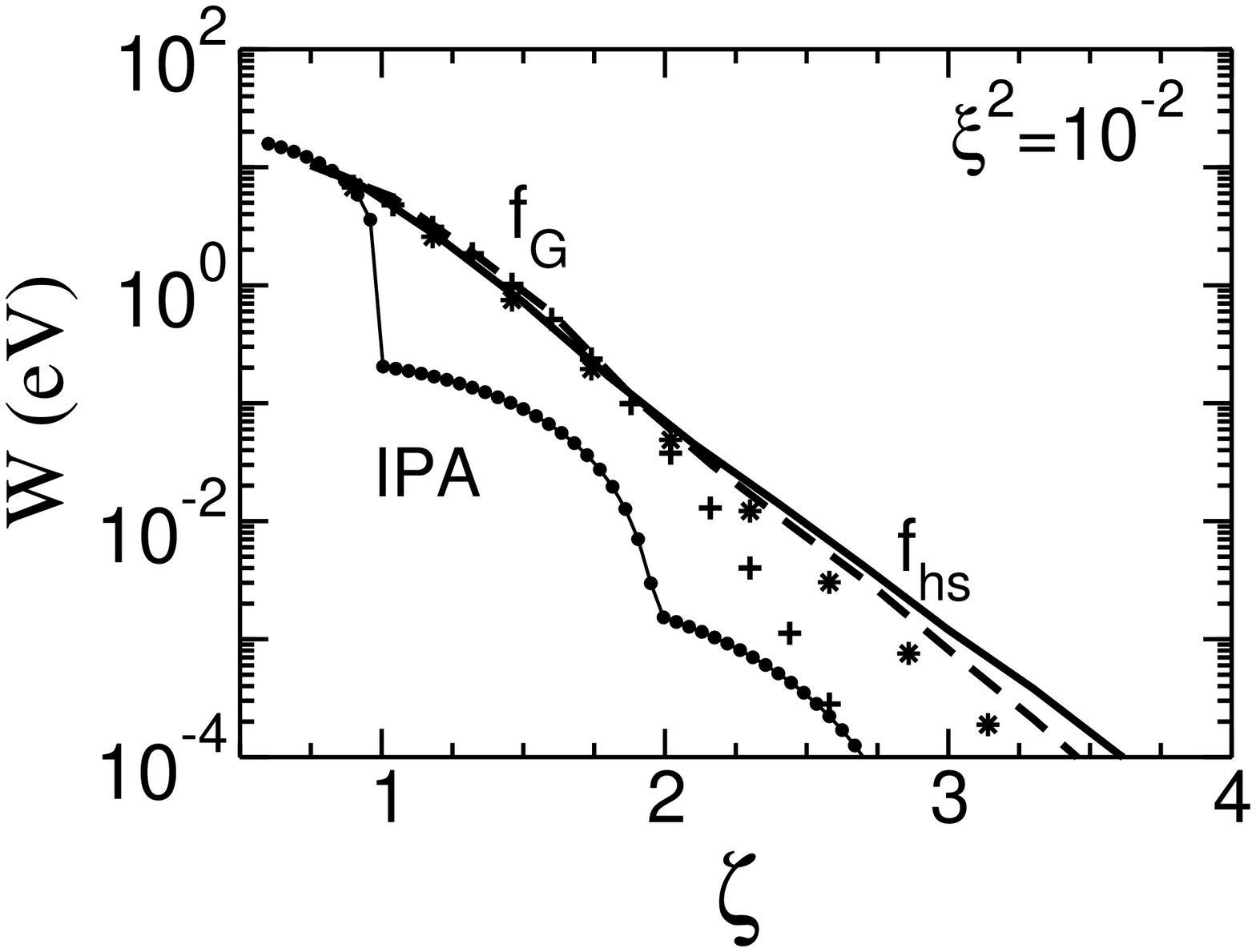}\qquad
\includegraphics[width=0.35\columnwidth]{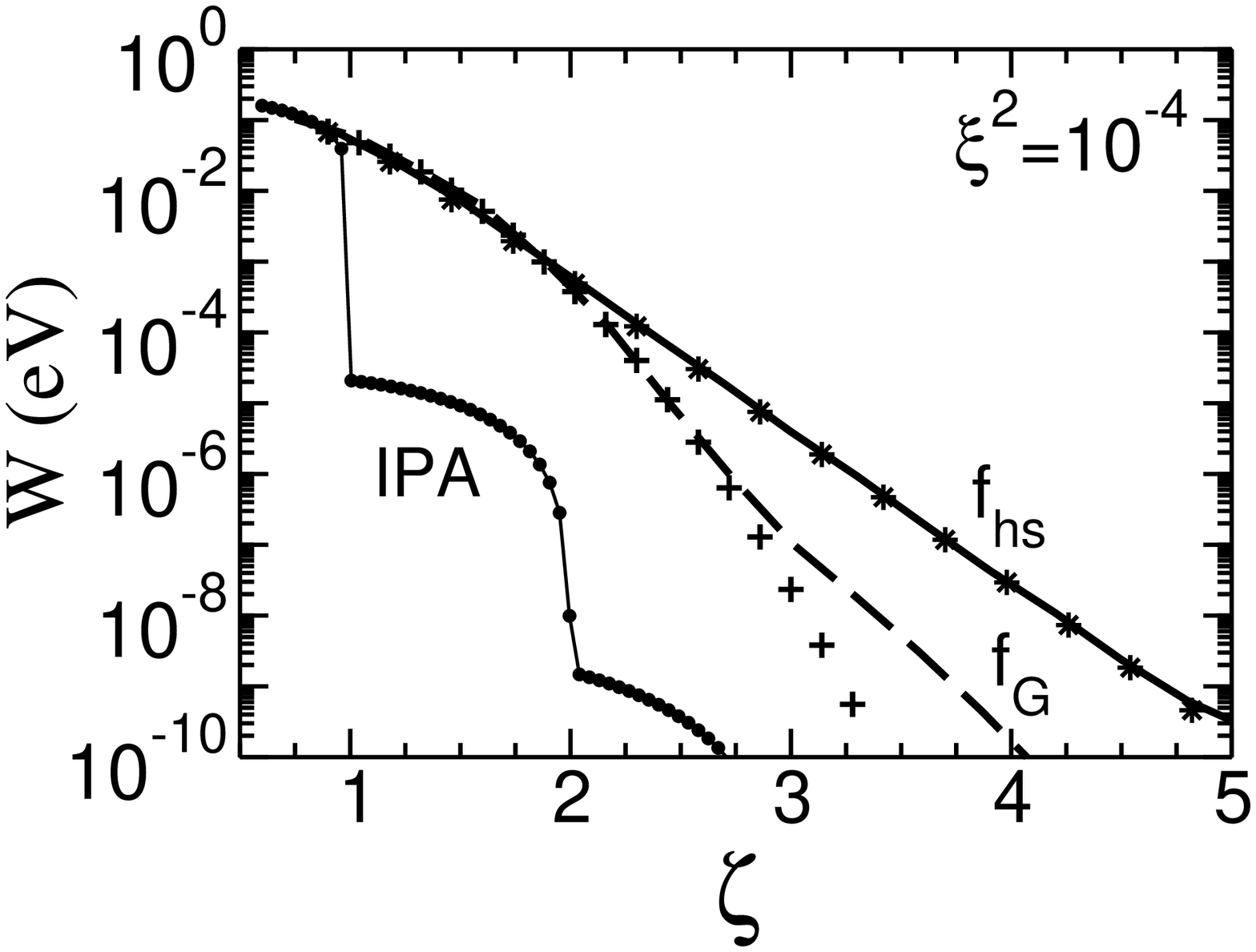}\\
\caption{\small{
The probability $W$ of $\ee$ production as a function
of the sub-threshold parameter $\zeta$ for one-parameter
envelope functions for  an ultra-short pulse with
$N=0.5$. The solid and dashed curves correspond
to numerical calculations with the hyperbolic secant and Gaussian
shapes, respectively. The symbols "star" and "plus"
are for the approximation~(\ref{U2}). The thin solid curves marked by
dots correspond to IPA. The left
and right panels are for $\xi^2=10^{-2}$ and  $10^{-4}$, respectively.
 \label{Fig:3} }}
\end{figure}

Our prediction for the total probability of $\ee$ pair production as
a function of the sub-threshold parameter $\zeta$ for the one-parameter
envelope functions for an ultra-short pulse with
$N=0.5$ is shown in Fig.~\ref{Fig:3}.
The solid and dashed curves exhibit results of   numerical calculations using Eq.~(\ref{III9})
with the hyperbolic secant and Gaussian
shapes, respectively.
The symbols "star" and "plus" display the resultsobtained by using
the approximation (\ref{U2}).
The thin solid curves marked by
dots correspond to the IPA case. The left
and right panels display results for $\xi^2=10^{-2}$ and  $10^{-4}$, respectively.
One can see an agreement of predictions for the ultra-short pulse
and IPA near and above the threshold at $\zeta\lesssim 1$, and a strong difference
between them below the threshold, i.e. for
$\zeta>1$. Our approximate (analytical)
solution of Eq.~(\ref{U2}) is in a fairly
well agreement with the complete numerical calculation.
The function $\Phi(l)$
in Eq.~(\ref{U2}) is rather smooth compared to the Fourier transform
$F(l-1)$, therefore, the dominant contribution to the integral
in  Eq.~(\ref{U2}) comes from the lower limit of $l$ and, qualitatively,
the slope of the probability as a function of $\zeta$ is determined
by the scale parameters of the envelope functions
\begin{eqnarray}
W_{\rm hs}(\zeta)\sim\exp\left[-\pi\Delta\zeta \right]~,\qquad
W_{\rm G}(\zeta)\sim\exp\left[-\tau^2_{\rm G}\zeta^2\right]~.
\label{U6}
\end{eqnarray}
Despite of the exponential decrease of the probability $W$
as a function of $\zeta$, one
can see a large difference (several orders of magnitude) between
predictions for the ultra-short pulse
and IPA (or "crossed field approximation"). In the latter case the probability
decreases much faster with increasing $\zeta$.

\begin{figure}[h!]
\includegraphics[width=0.35\columnwidth]{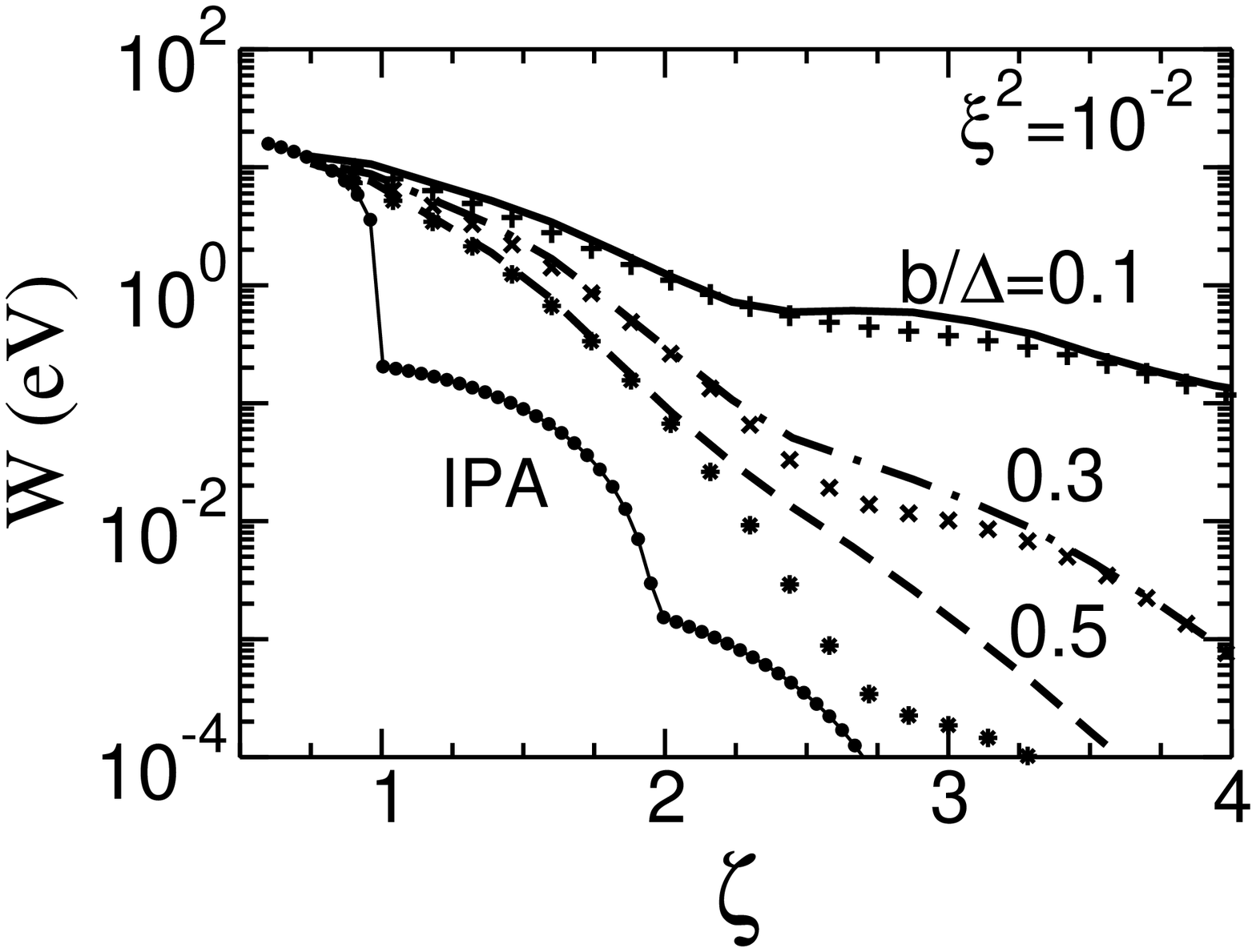}\qquad
\includegraphics[width=0.35\columnwidth]{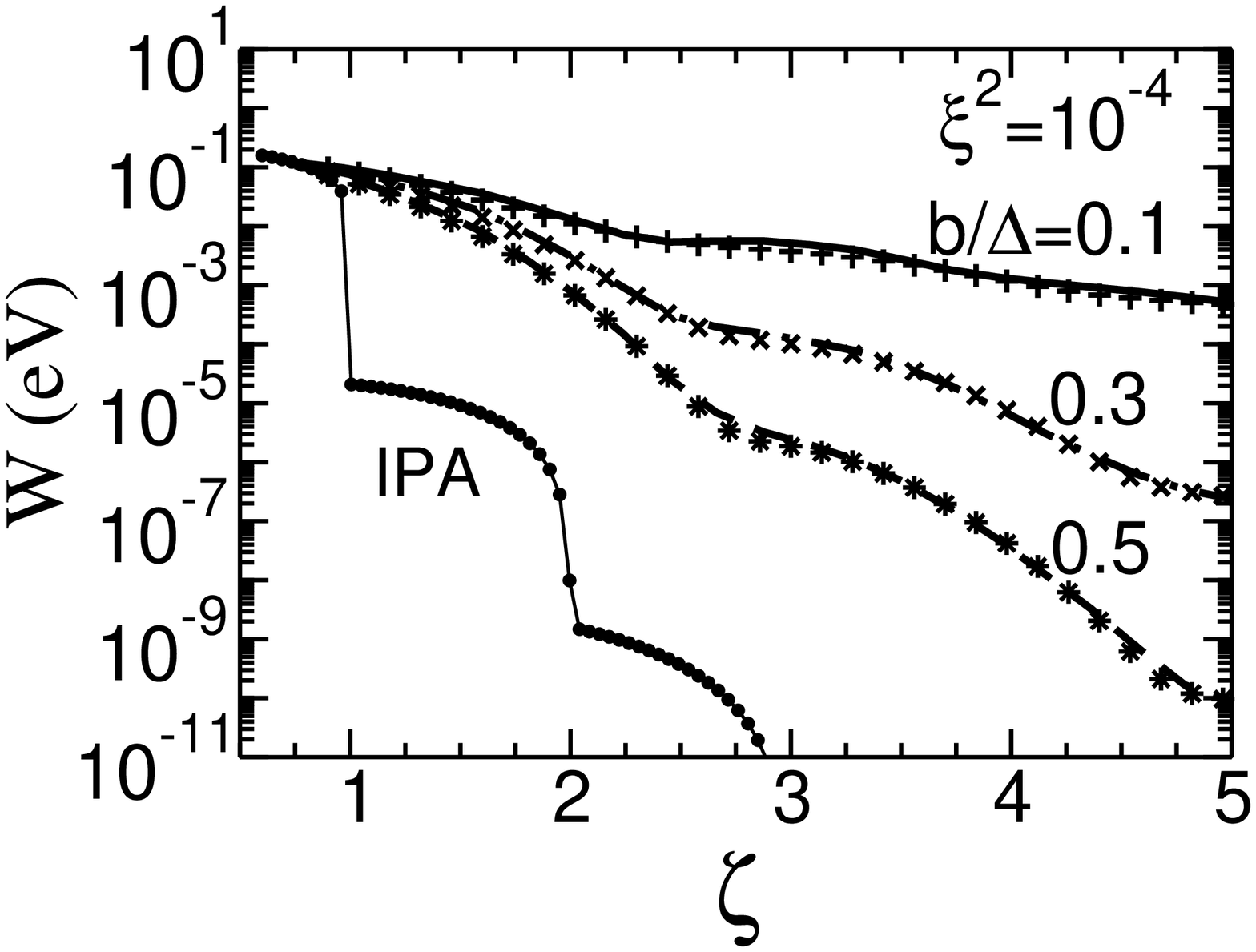}\\
\caption{\small{
The same as in Fig.~\ref{Fig:3} but for symmetrized Fermi shape envelope.
The solid, dot-dashed and dashed are for $b/\Delta=0.1$, 0.3
and 0.5, respectively.
The corresponding approximate solutions are shown by symbols "+", "x"
and "$\ast$", respectively.
 \label{Fig:4} }}
\end{figure}

Our results for the symmetrized Fermi envelope is presented in Fig.~\ref{Fig:4}.
Now, the shape of the probability is determined
by the two parameters $b$ (or $b/\Delta$)
and $\Delta$
\begin{eqnarray}
W_{\rm sf}(\zeta)\sim\exp\left[-2\pi\Delta\frac{b}{\Delta}\zeta \right]
\sin^2\Delta\zeta~.
\label{U7}
\end{eqnarray}
The first term describes the slope of the probability as a function of
$\zeta$.
The slope is proportional to the "ramping time"
of the envelope function, $b$ (or to the ratio $b/Delta$ at fixed $\Delta$).
The second term, following from the Fourier transform shown in Fig.~\ref{Fig:2},
describes some oscillations with a period inversely proportional
to the duration $\Delta$ of the flat-top section;
it is independent of
the ramping parameter $b$.
Again, one can see a great difference between predictions
for the ultra-short pulse and IPA on qualitative and quantitative levels.
The probability in IPA has a typical step-like behavior, where each new
step indicates the contribution of the next integer harmonic.
In FPA, the probability decreases monotonically with a slope
determined by the shape of the envelope.
The quantitative difference is rather large and,
as predicted by results shown in Figs.~\ref{Fig:3} and \ref{Fig:4}, can reach
 orders of magnitude depending on the shape of the envelope(s).

\subsection{Anisotropy}

As we have shown above, at small values of $z$, $z\ll1$,
the probability of $\ee$ production
is essentially determined by the pulse shape. The function $g(\phi)$ in
Eq.~(\ref{U01}) is not important and, therefore,
the total probability would be isotropic with respect
to the azimuthal angle $\phi_{e^-}=\phi_0$
because only the function ${\cal P}(\phi)$
contains a $\phi_0$-dependence.
For finite values of $z$,
the function $g(\phi)$ becomes important, and the electron
(positron) azimuthal angle distribution is anisotropic relative
to the direction of the vector $\vec a_x$ in Eq.~(\ref{III1}), at least
for the monotonically rapidly decreasing one-parameter
envelope shapes. The reason of an
anisotropy is the following. At finite values of $z$, the function
${Y}(l)$ in Eq.~(\ref{U01}) is determined by the integral over $d\phi$
with a rapidly oscillating function proportional to the exponential
\begin{eqnarray}
{\rm e}^{i
\left[ l\phi
 -z\left(\cos\phi_0\int\limits_{-\infty}^{\phi} d\phi'\, f(\phi')\cos\phi'
 + \sin\phi_0\int\limits_{-\infty}^{\phi} d\phi'\, f(\phi')\sin\phi'
 \right) \right]
 }~.
\label{U8}
\end{eqnarray}
In case of a fast-decreasing function $f(\phi')$,
the contribution of the term proportional to $\sin\phi_0$ is much smaller
compared to the term proportional to  $\cos\phi_0$.
At finite $z$,
the dominant contribution to the functions ${\cal Y}(l)$
comes from the region where
the difference in the exponent is minimal, i.e. $\phi_e=\phi_0 \simeq 0$.
This means that the electrons would be emitted mostly along
the vector $\vec a_x$
and the positrons in the opposite direction.

We define the anisotropy of the electron emission by
\begin{eqnarray}
{\cal A}=\frac{dW(\phi_e) - dW(\phi_e+\pi)}{dW(\phi_e) + dW(\phi_e+\pi)}~.
\label{U9}
\end{eqnarray}
\begin{figure}[h!]
\includegraphics[width=0.35\columnwidth]{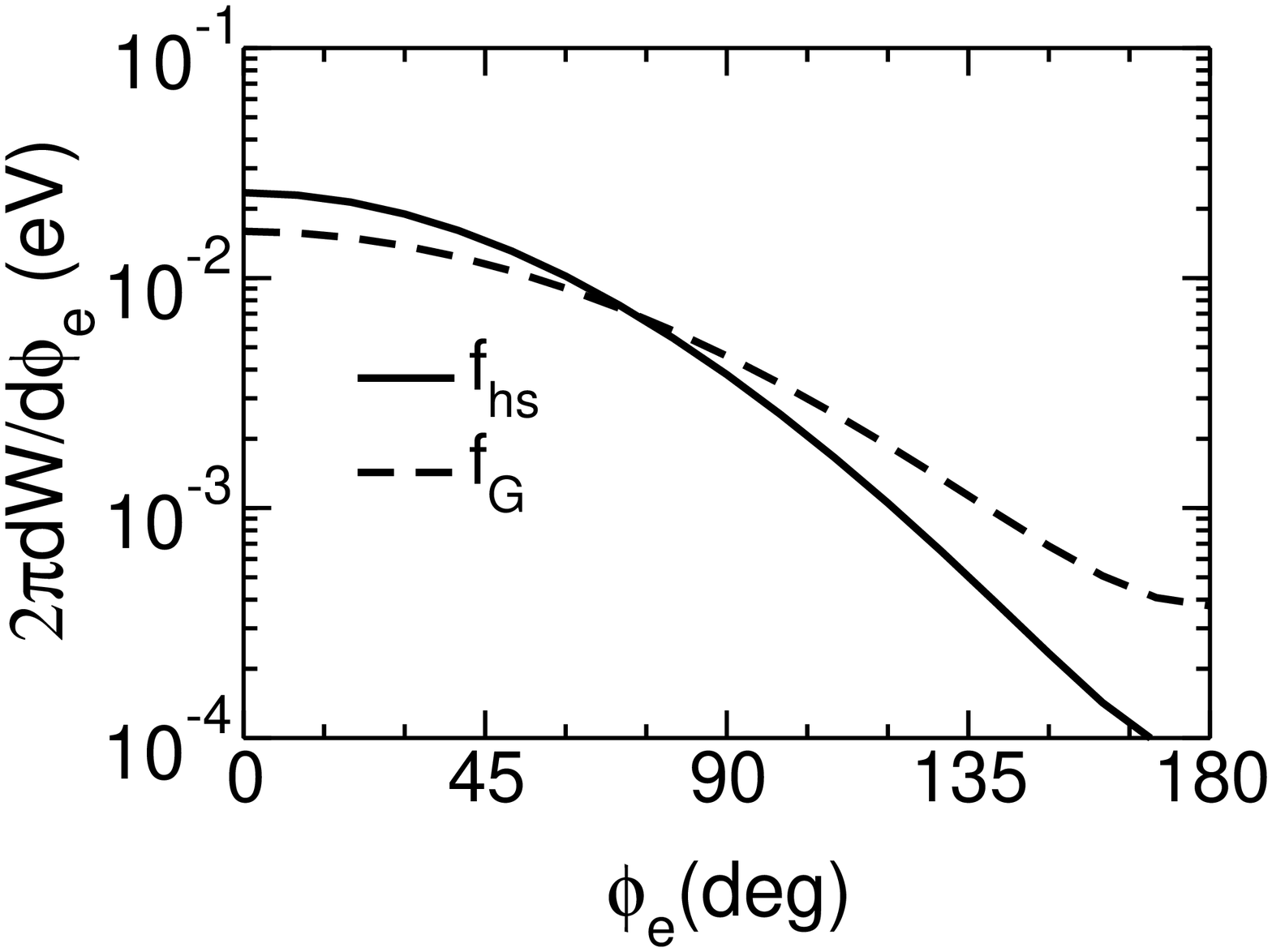}\qquad
\includegraphics[width=0.35\columnwidth]{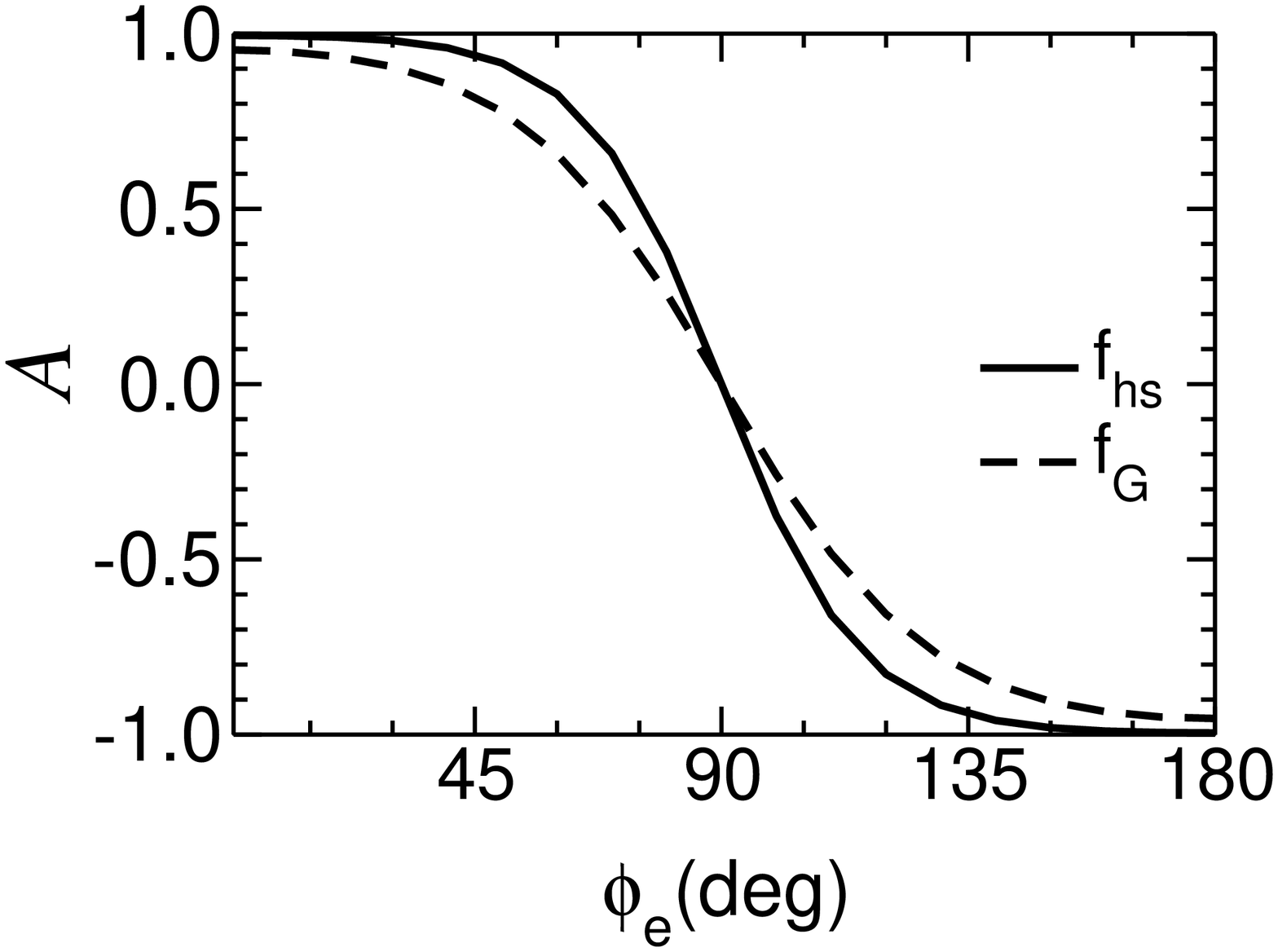}
\caption{\small{
Left panel: The differential production probability as a function of the azimuthal
angle $\phi_e$ of the electron emission.
Right panel: The anisotropy~(\ref{U9}) for  the hyperbolic secant
(solid curves) and Gaussian (dashed curves)
shapes. For $\xi^2=0.1$ and $ \zeta=4$.
 \label{Fig:5} }}
\end{figure}

The differential probability of the $\ee$ pair emission
and the anisotropy as a function of the azimuthal
angle $\phi_e$ are exhibited in Fig.~\ref{Fig:5}.
The calculations are for the fast-decreasing one-parameter
envelope functions for $\Delta=0.5\pi$,
$\zeta=4$ and $\xi^2=0.1$. One can see
a rapidly decreasing probability with $\phi_e$ which leads to the strong
anisotropy  of electron (positron) emission.

In case of the symmetrized Fermi distribution with
small $b/\Delta$, the situation changes drastically.
As $b/\Delta\to 0$ the envelope function goes to
the flat-top (step-like)
shape $f_{\rm Fs}(\phi)\to \theta(\Delta^2-\phi^2)$ with $\theta(x)=1$, 0
for $x\ge 0$ or $x< 0$, respectively, and correspondingly
\begin{eqnarray}
{\cal Y}(l)=\frac{1}{2\pi}
\int\limits_{-\Delta}^{\Delta}
d\phi \, {\rm e}^{i\left[\tilde l\phi -z\sin(\phi-\phi_0) \right]}
\label{U99}
\end{eqnarray}
 with $\tilde l= l+ \xi^2\zeta u$. The function ${\cal Y}(l)$
 in the  region $\zeta\leq l< l_{\max}\gg1$
 is alternating, rapidly oscillating with an amplitude that depends
 only on $\xi$, $\zeta$, and $u$.
 It is not sensitive to $\phi_0$.  Modifications of
 $\phi_0$ lead some phase shift of ${\cal Y}(l)$ in a range of integration,
 leaving $\langle|{\cal Y}(l)|^2\rangle$ to be independent of $\phi_0$
 \begin{figure}[h!]
\includegraphics[width=0.35\columnwidth]{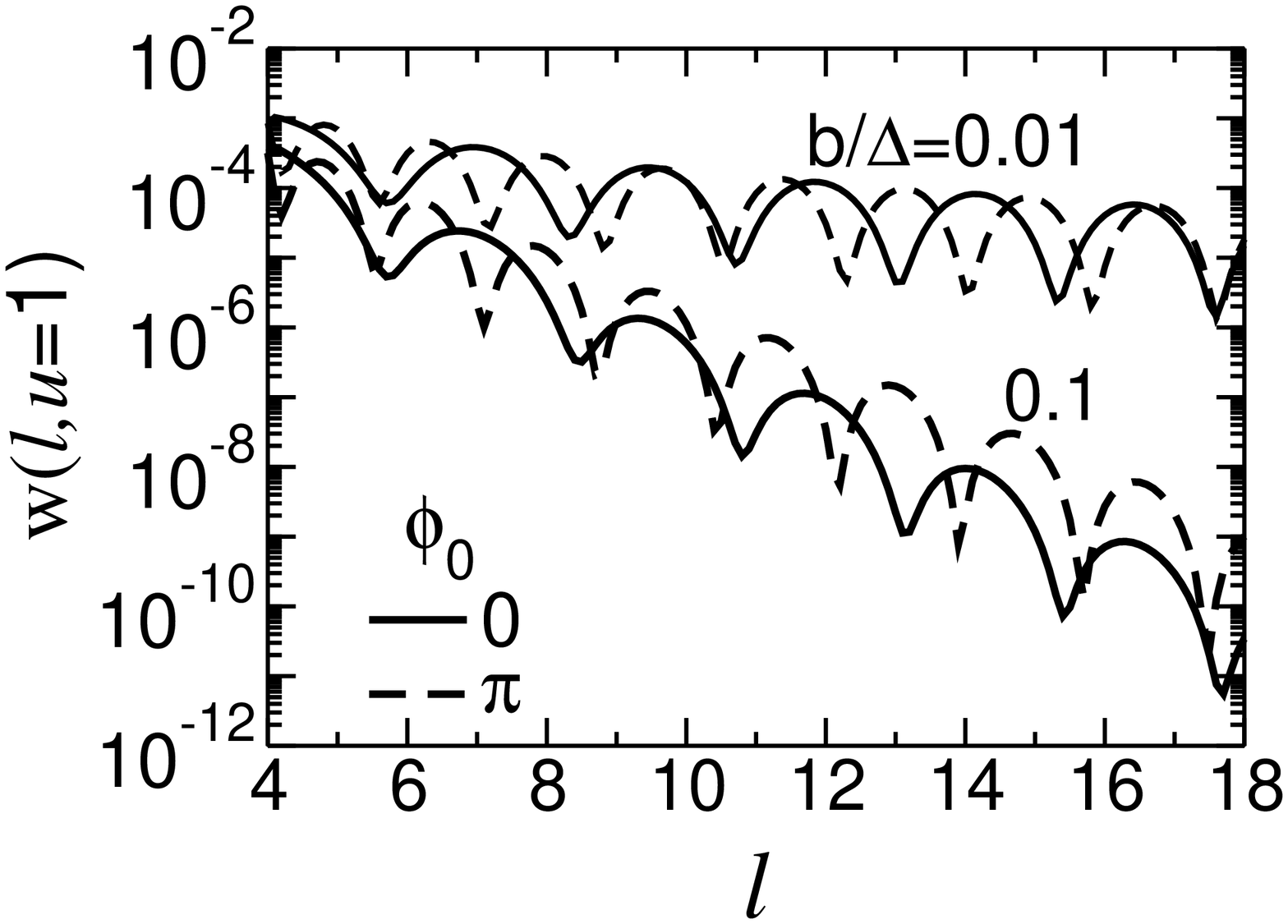}\qquad
\includegraphics[width=0.35\columnwidth]{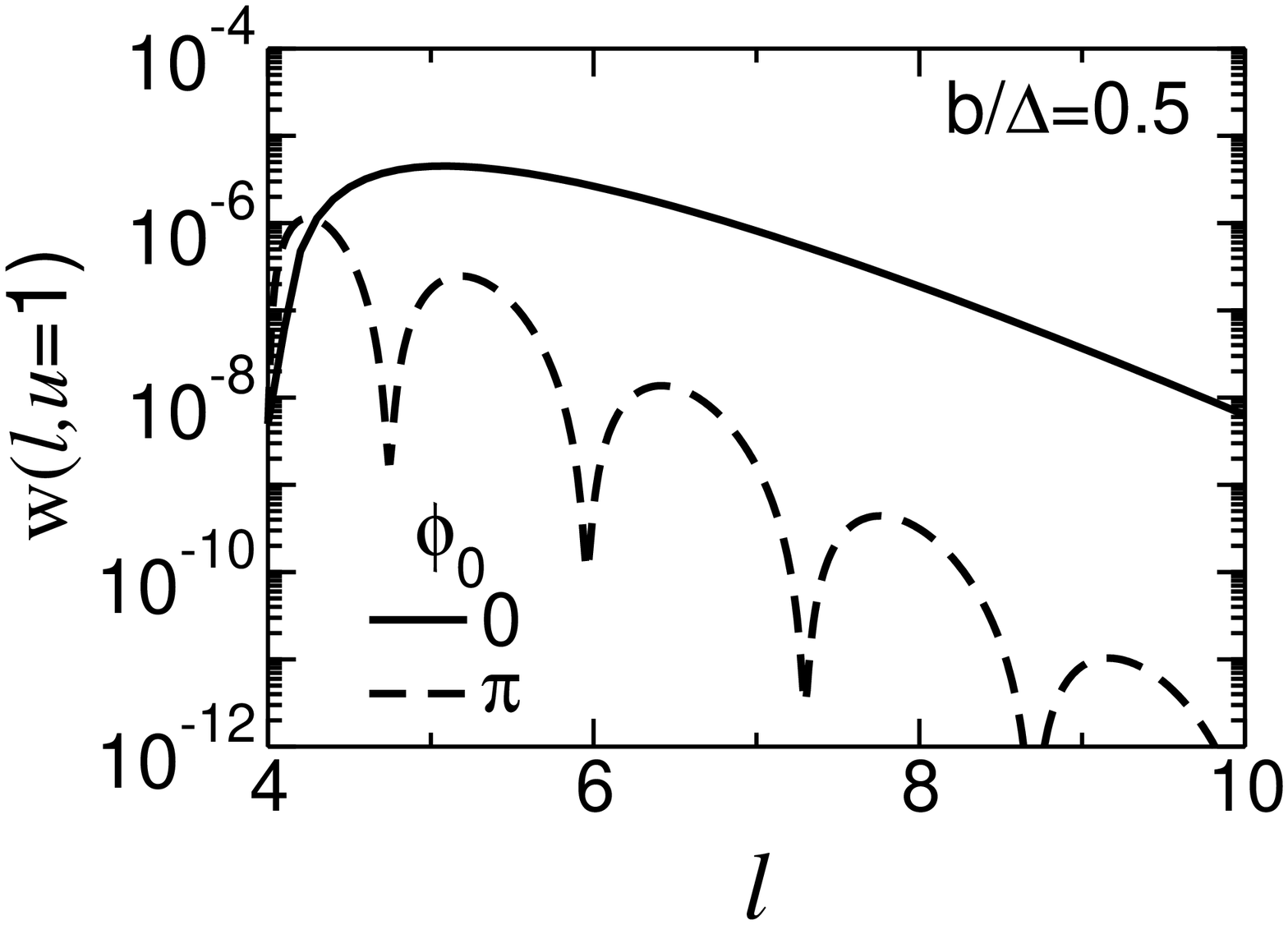}
\caption{\small{
The partial probability $w(l)$ defined in~(\ref{III9})
at $\phi_0=0$ and $\pi$ shown by solid and
dashed curves, respectively, for the symmetrized Fermi envelope shape.
The left panels correspond to small values of
$b/\Delta=$0.01 and 0.1, while the right panel is for $b/\Delta=0.5$.
For $\xi^2=0.1$ and $ \zeta=4$.
 \label{Fig:6} }}
\end{figure}
Therefore, the dependence of the integral of
the partial probability $w(l)\sim|{\cal Y}(l)|^2$ in Eq.~(\ref{III9})
on $\phi_0$ is negligible.
As an example, in the left panel of Fig.~\ref{Fig:6} we present the
partial probability $w(l)$ as a function of $l$, calculated at
$\xi^2=0.1$, $\zeta=4$ and $u=1$ for the small values of
$b/\Delta$ equal to 0.1 and 0.01 at $\phi_0=0$ and $\pi$, shown by
solid and dashed curves, respectively. One can see some small
modification of the frequency of oscillations at $l\sim l_{\rm min}=\zeta$
at two extreme values of $\phi_0$, but the amplitudes of the oscillations
are similar. This situation is quite different from the case of the large
value of $b/\Delta=0.5$ presented in the right panel of Fig.~\ref{Fig:6}.
One can see a strong difference in the $l$-dependence of $w(l)$ for $\phi_0=0$
and $\pi$. In the first case, the function
$w(l)$ has only one oscillation in a wide range of $l$ and decreases smoothly
with $l$. In the second case, the probability has a number of oscillations
decreasing rapidly with increasing $l$. As a result, the total probability
in the second case is much smaller.

This behavior can also be understood from a different point of view.
 The integral over $l$ of the derivative
 of the partial probability $w(l)$ in Eq.~(\ref{III9})
 \begin{eqnarray}
 \frac{dw(l)}{d\phi_0}
 \sim \frac{d}{d\phi_0}|{\cal Y}(l|^2
 \sim |{\cal Y}(l\pm 1){\cal Y}(l)|
\label{U10}
\end{eqnarray}
 is vanishing because of the alternating and oscillating nature of
 ${\cal Y}(l)$. Therefore, the probability $W$ is independent
 of $\phi_0$.

\begin{figure}[h!]
\includegraphics[width=0.35\columnwidth]{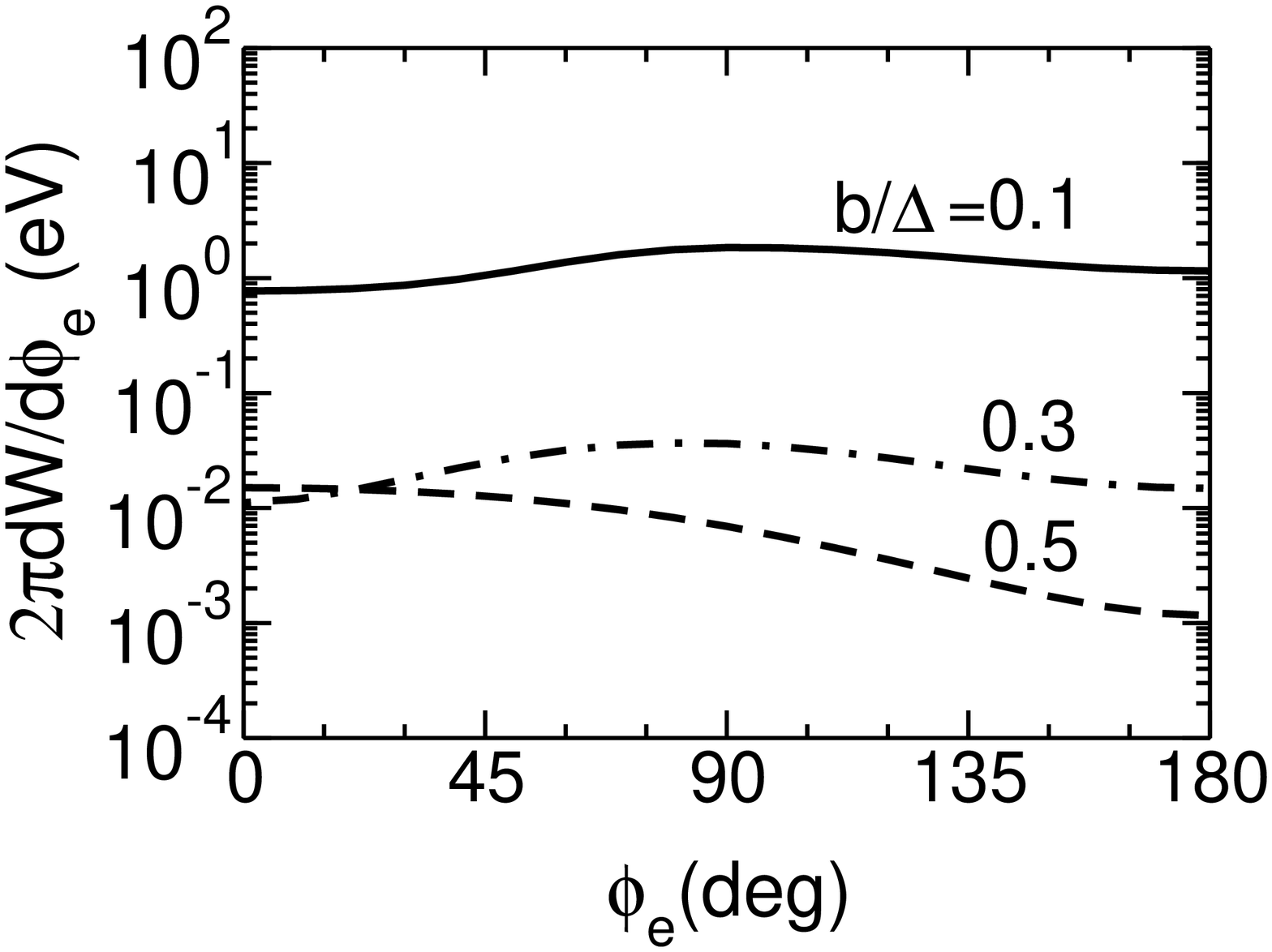}\qquad
\includegraphics[width=0.35\columnwidth]{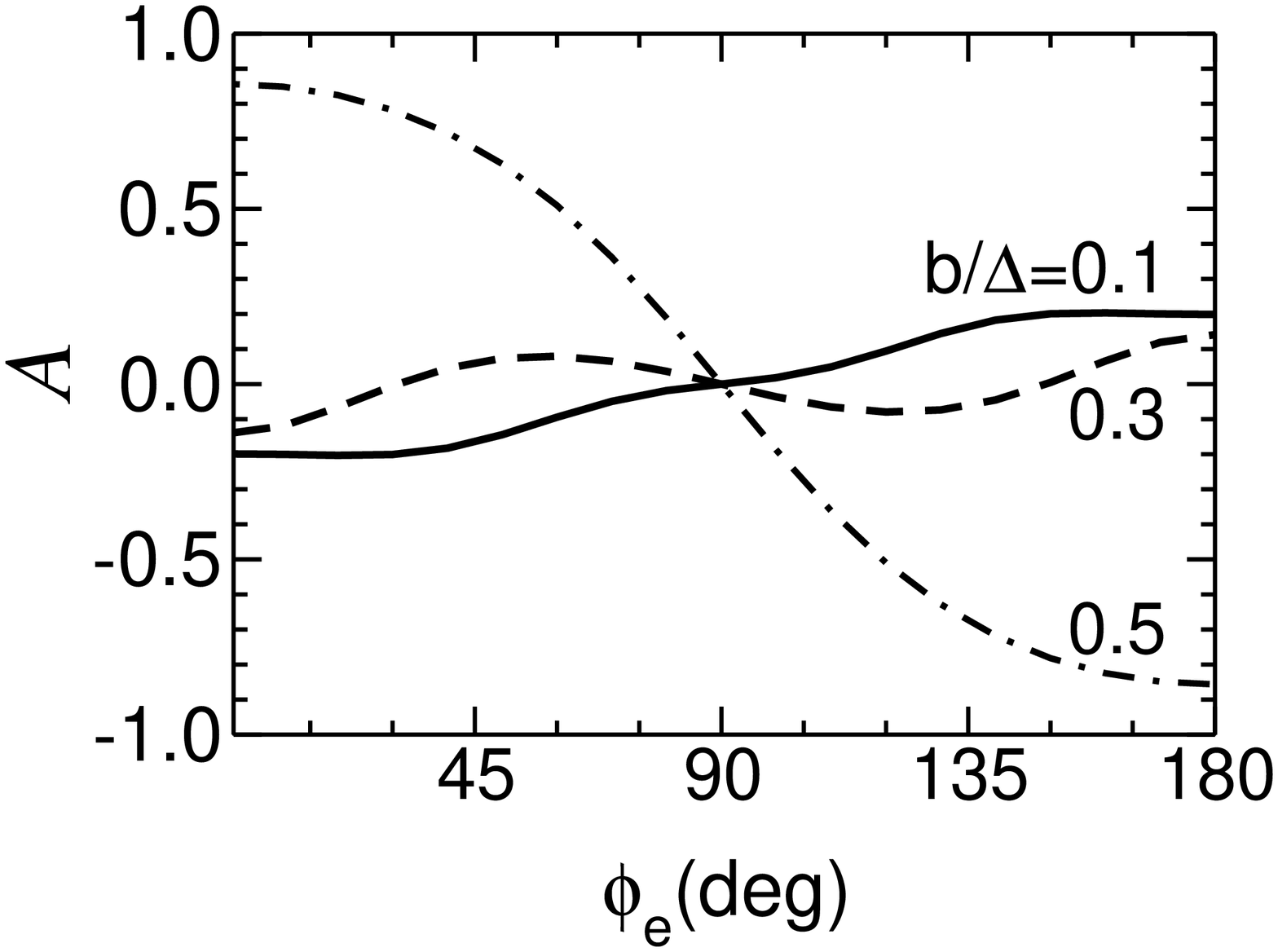}
\caption{\small{
The same as in  Fig.~\ref{Fig:5} but for the symmetrized Fermi shape.
The solid, dot-dashed and dashed curves are for
$b/\Delta=0.1$, 0.3 and 0.5, respectively. For $\xi^2=0.1$
and $ \zeta=4$.
 \label{Fig:7} }}
\end{figure}

In Fig.~\ref{Fig:7} we present our results for
the symmetrized Fermi shape for the production probability (left panel)
and for the anisotropy (right panel) for $b/\Delta=0.1$, 0.3 and 0.5.
The result for $b/\Delta=0.5$ is similar to that
shown in Fig.~\ref{Fig:5}. However, for smaller values of $b/\Delta$,
the probability is a smooth function of $\phi_0$ with some modest
enhancement around $\pi/2$, which leads to a negligible anisotropy.


\section{Short pulses}

In this section
we  consider short pulses with the number of oscillation
$N\geq 2$, however, many results are valid even for pulses with
$N\sim1$.
As we have seen, the one-parameter envelope shapes lead to similar results
even for ultra-short pulses, therefore, later on we will limit our consideration
to two extreme envelope shapes:
hyperbolic secant and symmetrized Fermi
shape with $b/\Delta=0.1$.

\begin{figure}[h!]
\includegraphics[width=0.3\columnwidth]{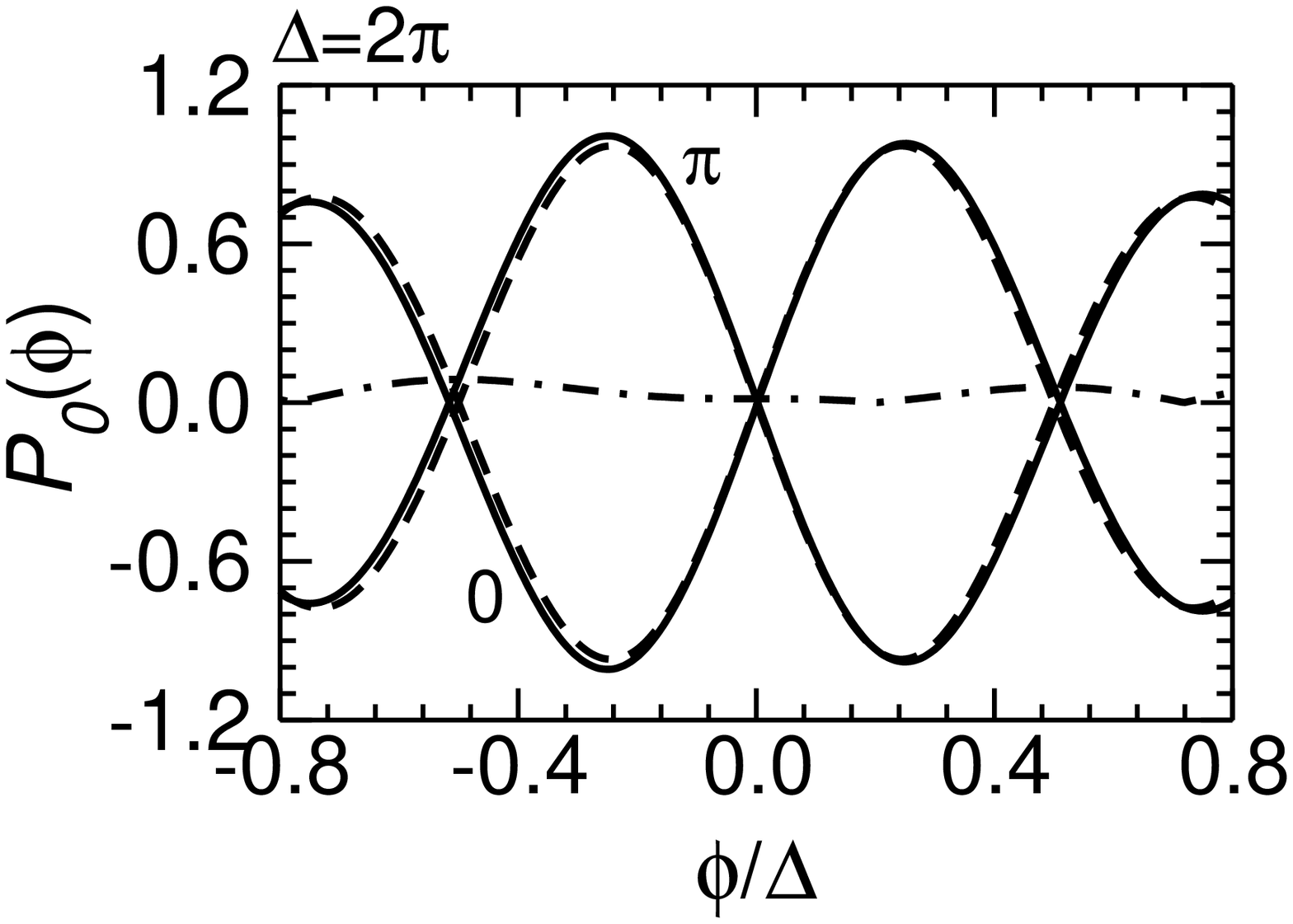}\qquad
\includegraphics[width=0.3\columnwidth]{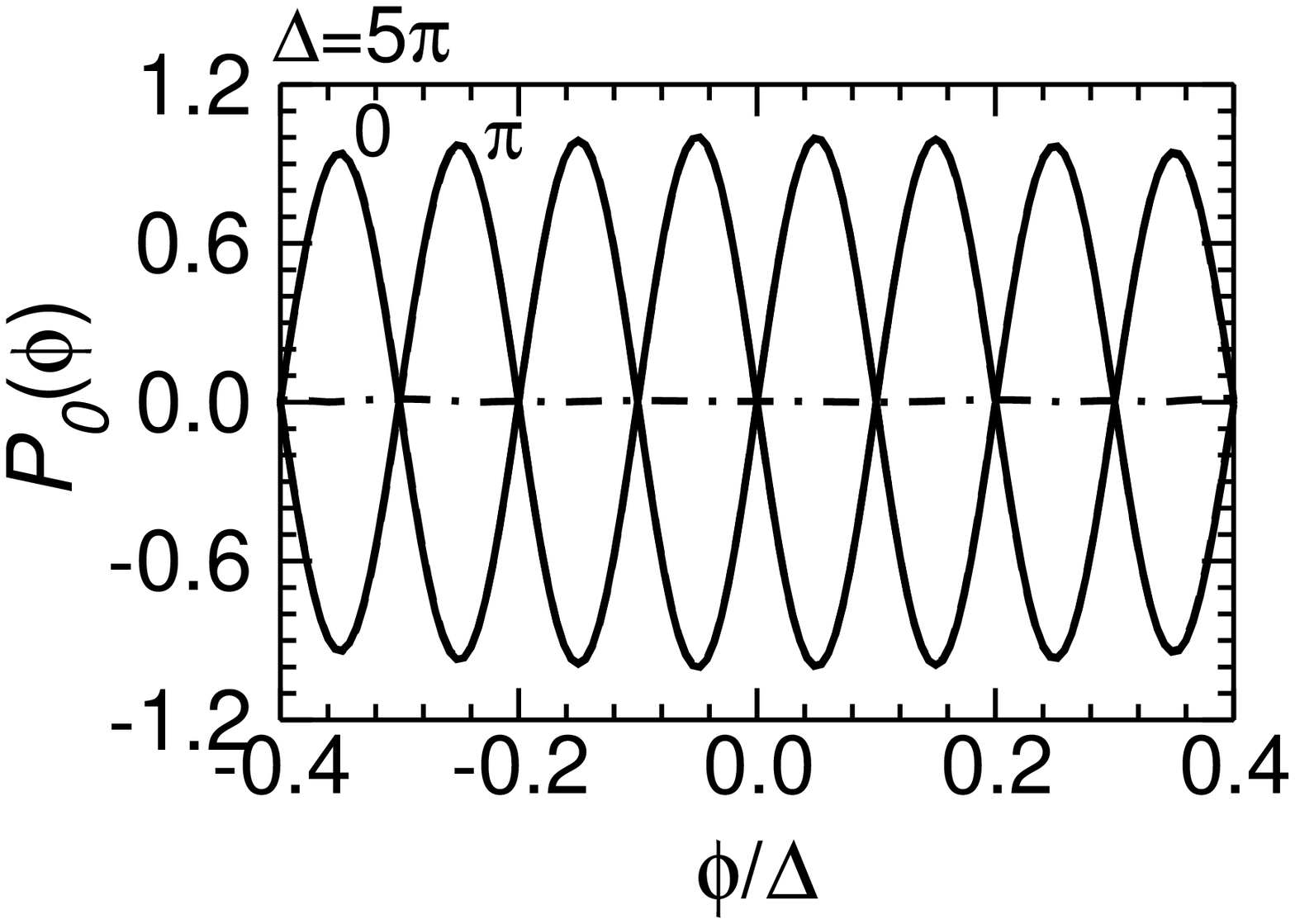}
\caption{\small{
The function ${\cal P}_0(\phi)$ defined in (\ref{III21})
with (solid curves) and without (dashed curves)
the term~(\ref{III22}) for $\Delta=\pi\,N$ with
$N=2$ and 5, shown in left and right panels, respectively.
The term~(\ref{III22}) is shown separately
by dot-dashed curves.
\label{Fig:8} }}
\end{figure}

As mentioned above,  Eqs.~(\ref{III9}) and~(\ref{III20}) can be
used for numerical estimates of the $\ee$ production probability
evaluating five dimensional integral(s) with rapidly oscillating
functions. Technically, such an approach needs long calculation
time which makes it difficult for applications in transport/Monte
Carlo codes. However, a closer inspection of the functions ${\cal
P(\phi)}$ and $C^{(i)}(l)$ shows that the number of integrations
may be reduced and, in some cases, Eq.~(\ref{III20}) may be
expressed in an analytical form. Thus, integrating by parts the
function $\cal P(\phi)$ might be rewritten in the following form
\begin{eqnarray}
{\cal P}(\phi)\equiv{\cal P}_0(\phi)-
\xi^2\zeta u\int\limits_{-\infty}^\phi d\phi'\,f^2(\phi'),\qquad
{\cal P}_0(\phi)
=z\,\left(\sin(\phi-\phi_0)f(\phi)
+ {\cal O}\left(\frac{1}{\Delta}\right) \right)\,
\label{III21}
\end{eqnarray}
with
\begin{eqnarray}
{\cal O}\left(\frac{1}{\Delta}\right)=
-\frac{1}{\Delta}\int\limits_{-\infty}^\phi\,d\phi'\,
\sin(\phi'-\phi_0)f'(\phi')~.
\label{III22}
\end{eqnarray}
The contribution of this term to $\cal P(\phi)$ is sub leading
for the finite pulse size $\Delta=\pi N$ with $N\ge2$.
First, because of the explicit factor ${1}/{\Delta}$, and second
because the derivative $f'(\phi)$ in the integrand reaches
its maximum value at the boundaries of the pulse,
where this function is suppressed.
For an illustration, in Fig.~\ref{Fig:8}
we present results of a numerical analysis of ${\cal P}_0(\phi)$
with the hyperbolic secant envelope function.
The solid and dashed curves exhibit calculations with and without
the term~(\ref{III22}), respectively for $\phi_0=0$ and $\pi$.
The left and right panels correspond to
$\Delta=\pi\,N$ with $N=2$ and 5, respectively.
The term ($|{\cal O}({1}/{\Delta})|$) is shown
by dot-dashed curves.
One can see, in fact, that this term is rather small and may be
omitted. For the  flat-top envelopes this approximation
is even much better.

Using this approximation one can
express the basic functions $C^{(i)}(l)$
defined in Eqs.~(\ref{III5}) and (\ref{III8})
through the new functions
$Y_l$ and $X_l$, which may be considered as an analog
of the Bessel functions in IPA,
\begin{eqnarray}
Y_l(z)&=&\frac{1}{2\pi}
\int\limits_{-\infty}^{\infty}\, d\psi\,\tilde{f}(\psi + \phi_0)
\,{\rm e}^{il\psi-iz\sin\psi} ~,\nonumber\\
X_l(z)&=&\frac{1}{2\pi}
\int\limits_{-\infty}^{\infty}\, d\psi\,\tilde{f^2}(\psi + \phi_0)
\,{\rm e}^{il\psi-iz\sin\psi}~,\nonumber\\
\tilde{f}(\phi)&=&f(\phi)\,\exp[i\xi^2\zeta\,u\,r(\phi)]~,\qquad
\tilde{f^2}(\phi)=f^2(\phi)\,\exp[i\xi^2\zeta\,u\,r(\phi)]~,\nonumber\\
r(\phi)&=&\int\limits_{-\infty}^\phi  d\phi'\, f^2(\phi')~,
\label{III24}
\end{eqnarray}
where the function $r(\phi)$ is a smooth function of $\phi$. For
the hyperbolic  secant we have $r(\phi)=\Delta
\tanh(\phi/\Delta)$, where we skip the constant term which does
not contribute; for the flat-top envelope,
$r(\phi)\sim\phi\,\theta(\Delta^2 -\phi^2)$. The new
representation of the basic functions $C^{(i)}(l)$ reads
\begin{eqnarray}
C^{(1)}(l)&=&X_l(z)\,{\rm e}^{i(l)\phi_0}~,\nonumber\\
C^{(2)}(l)&=&\frac{1}{2}\left( Y_{l+1}{\rm e}^{i(l+1)\phi_0}
+ Y_{l-1}{\rm e}^{i(l-1)\phi_0}\right)~,\nonumber\\
C^{(3)}(l)&=&\frac{1}{2i}\left( Y_{l+1}{\rm e}^{i(l+1)\phi_0}
- Y_{l-1}{\rm e}^{i(l-1)\phi_0}\right)~,\nonumber\\
C^{(0)}(l)&=&\widetilde Y_l(z){\rm e}^{i(l)\phi_0},
\qquad \widetilde Y_l(z)=\frac{z}{2l}
\left(Y_{l+1}(z) + Y_{l-1}(z)\right) - \xi^2\frac{u}{u_l}\,X_l(z)~.
\label{III25}
\end{eqnarray}
It allows to express
$w(l)$ in Eq.~(\ref{III20}) in the form
\begin{eqnarray}
w(l)=
 2 \widetilde Y^2_l(z) +\xi^2(2u-1)
\left(Y^2_{l-1}(z)+ Y^2_{l+1}(z)-2\widetilde Y_l(z)X^*_l(z)\right)~,
\label{III26-0}
\end{eqnarray}
which resembles the expression for the probability in case of IPA
(cf. Eq.~(\ref{II8-1})).
Now we are going to discuss separately the weak-, intermediate-
and strong-field regimes.

\subsection{Production probability  at small field intensities ($\xi^2\ll1$) }

In case of small values of $\xi^2$,
$\xi^2\ll1$, implying  $z<1$, we decompose $l=n +\epsilon$,
where $n$ is the integer part of $l$,
yielding
\begin{eqnarray}
Y_l&\simeq&\frac{1}{2\pi}\int\limits_{-\infty}^{\infty}\,d\psi
\,{\rm e}^{il\psi -iz\sin\psi} f(\psi+\phi_0)\nonumber\\
&=&
\frac{1}{2\pi}\int\limits_{-\infty}^{\infty}\,d\psi
\sum\limits_{m=0}^{\infty}
\frac{(iz)^m}{m!}\sin^m\psi
\,{\rm e}^{i(n+\epsilon)\psi} f^{m+1}(\psi+\phi_0)~.
\label{B5}
\end{eqnarray}
Similarly, for the function $X_l(z)$ the
substitution $f^{m+1}\to f^{m+2}$ applies.
The dominant contribution to the integral with rapidly oscillating
integrand comes from the term with $m=n$, which result in
\begin{eqnarray}
Y_{n+\epsilon}\simeq
\frac{z^n}{2^nn!}\,{\rm e}^{-i\epsilon\phi_0}
F^{(n+1)}(\epsilon)~,\qquad
X_{n+\epsilon}\simeq
\frac{z^n}{2^nn!}\,{\rm e}^{-i\epsilon\phi_0}
F^{(n+2)}(\epsilon)~,
\label{B6}
\end{eqnarray}
where the function $F^{(n)}(\epsilon)$ is the Fourier transform of the function
$f^n(\psi)$.

As an example, let us analyze the
$\ee$ production near the threshold, i.e. $\zeta\sim1$.
In this case, the contribution with
$n=1$ is dominant and, therefore,
the functions  $Y_{0+\epsilon}$ are crucial, including the first term
in (\ref{III26-0}). The functions
$X_{0 +\epsilon}$ are not important because they are
multiplied  by the small $\xi^2$
and may be omitted.
Negative $\epsilon=\zeta-1$ and positive
$\epsilon$ correspond to the above- and sub-threshold pair production,
respectively. The function $Y_{0+\epsilon}$ reads
$Y_{0+\epsilon}=F^{(1)}(\epsilon)\,\exp[-i\phi_0\epsilon]$
with corresponding Fourier transform
$F^{(1)}(\epsilon)$ presented in Eq.~(\ref{U5}).
Note that the $\phi_0$-dependence of the production probability
disappears in this case because the latter one is determined by the
quadratic terms of the $Y$-function.

Consider first the pair production above the threshold.
Keeping the terms with leading power of $\xi^2$ one can
express the  production probability
as
\begin{eqnarray}
\frac{d W}{du}=\frac{\alpha M_e\zeta^{1/2}}{4N_0}
\,\left[ \frac{u}{u_1}\left(1-\frac{u}{u_1}\right) +u  - \frac12
\right]
\frac{\xi^2}{u^{3/2}\sqrt{(u-1)}}\, I_0~,
\label{III28}
\end{eqnarray}
where, taking into account that, at finite values of $\Delta$,
Fourier transforms for all considered envelopes
decrease rapidly with increasing $\epsilon$ one can get
\begin{eqnarray}
I_0
\simeq\int\limits_{1-\zeta}^{1/2} d\epsilon\, F^{(1)}{}^2(\epsilon)
\simeq\int\limits_{-\infty}^{\infty} d\epsilon\, F^{(1)}{}^2(\epsilon)
=\frac{1}{2\pi}\int\limits_{-\infty}^{\infty}
d\phi\,f^2(\phi)\simeq N_0~.
\label{III29}
\end{eqnarray}
Combining these two equations one recovers exactly the IPA
result~\cite{Ritus-79}. Thus, we can conclude that for small field
intensities for a finite pulse duration the probabilities of $\ee$
pair emission above threshold with $\zeta<1$ results in a
coincidence of IPA and FPA, independently of the shape of the
envelope function.
\begin{figure}[h!]
\includegraphics[width=0.35\columnwidth]{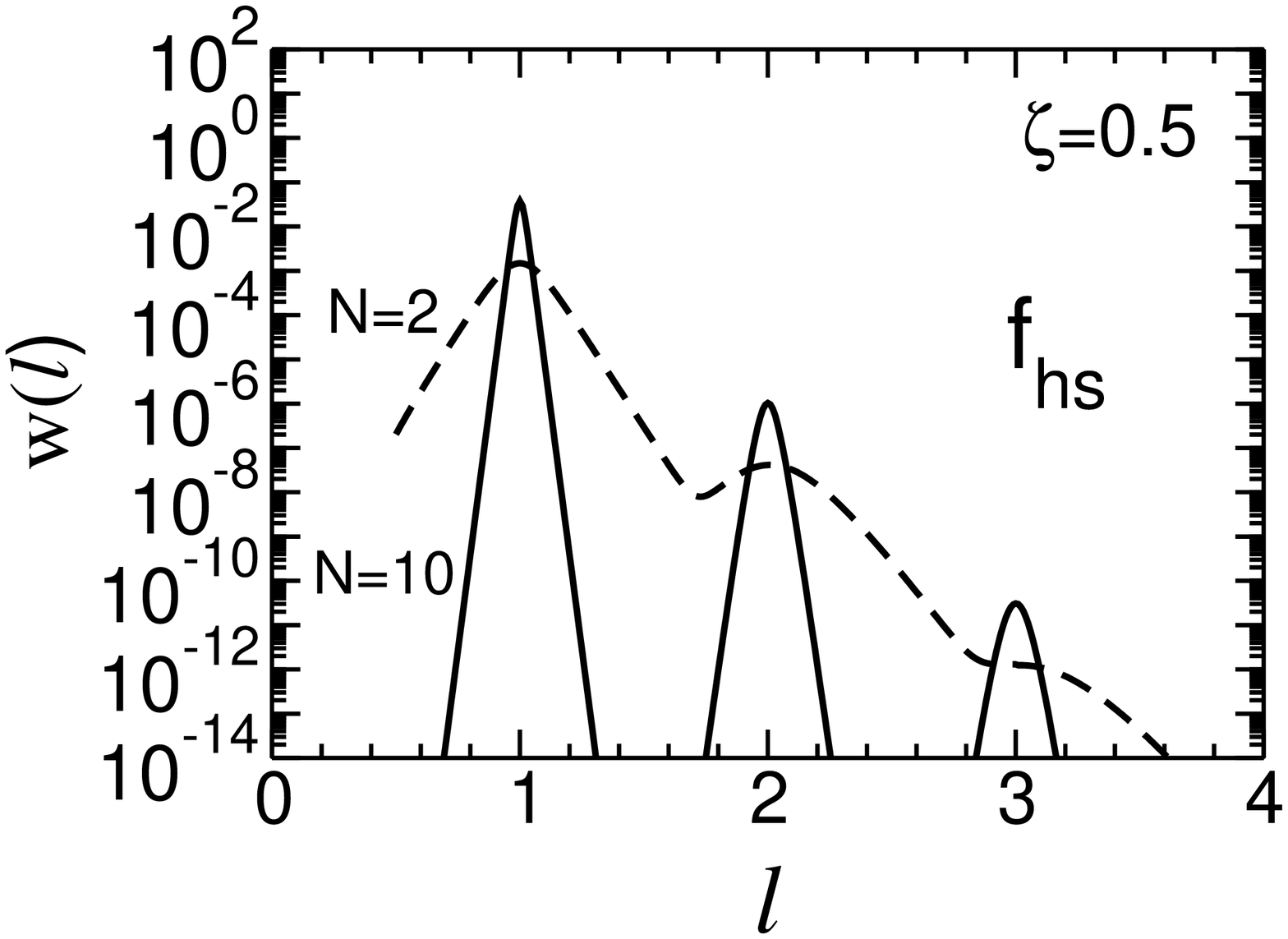}\qquad
\includegraphics[width=0.35\columnwidth]{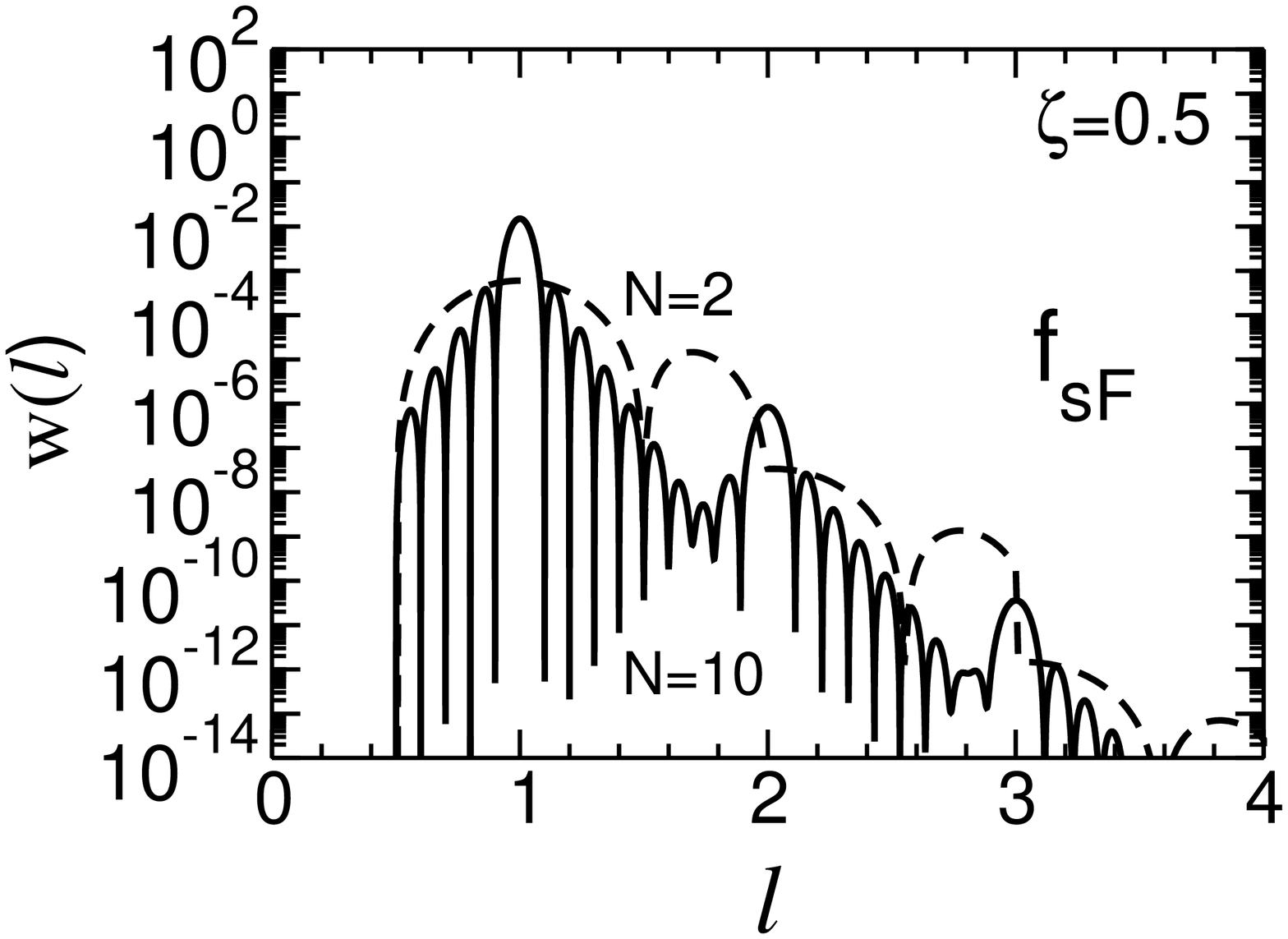}
\caption{\small{The partial probability  $w(l)$
defined in (\ref{III26-0}) as a function of $l$
at $u=1$. The solid and dashed curves correspond to the
beam size $\Delta=\pi N$ with $N=2$ and 10, respectively.
Left and right panels exhibit results for the envelopes
with hyperbolic secant and symmetrized Fermi shapes,
respectively. For $\xi=10^{-4}$ and $\zeta=0.5$
\label{Fig:9} }}
\end{figure}
For an illustration, in Fig.~\ref{Fig:9} we show the partial
probability $w(l)$, calculated at $u=1$  for the above-threshold
region with $\xi^2=10^{-2}$ and $\zeta=0.5$ in a finite region of
$l$ for the envelope size $\Delta=\pi N$ with $N=2$ and 10,
respectively. For the envelope with a hyperbolic secant shape
(left panel) one can see smooth curves with maxima at integer
values of $l$. The widths of bumps decrease with increasing $N$.
However, the integral of $w(l)$ over $l$ in the neighborhood of
the first maximum is independent of $N$ and coincides with the
contribution of the first harmonic in IPA which leads to an
equality of IPA and FPA results. For the symmetrized Fermi shape
(right panel) the situation is different in some sense. The
corresponding Fourier transforms
 $F^{(n)}_{sF}(l)$ in (\ref{B6})
oscillate with $l$. For example, the function $F^{(1)}_{sF}$
goes to zero at a multiple of $1/N$. This results
in an oscillating structure of $w(l)$. However, the exponential
decrease of $w(l)$ with increasing of the integer values of $l$
is the same.

The situation changes when we are slightly below threshold, i.e.
$\zeta>1$. In this case, the function $Y_{0+\epsilon}$ dominates
again and the result for FPA is the same as in (\ref{III28}) but
with the substitution $I_0\to I_1$, with $I_1
\simeq\int\limits_{\zeta-1}^{1} d\epsilon\,
F^{(1)}{}^2(\epsilon)$. In case of smooth envelope shape
(e.g.~hyperbolic secant) the dominating contribution to this
integral comes from the lower limit and, therefore, $I_1\sim
\left(F_{\rm hs}^{(1)}(\zeta -1)\right)^2$. As a result, the
production probability strongly depends on the duration $\Delta$
of the pulse.

In case of a flat-top envelope, we have a similar effect, because
$F_{sF}^{(1)}(l)$ in general decreases
exponentially as $\exp(-\pi b l)$, where
$b$ increases with increasing $N$ at fixed $b/\Delta$.

\begin{figure}[h!]
\includegraphics[width=0.35\columnwidth]{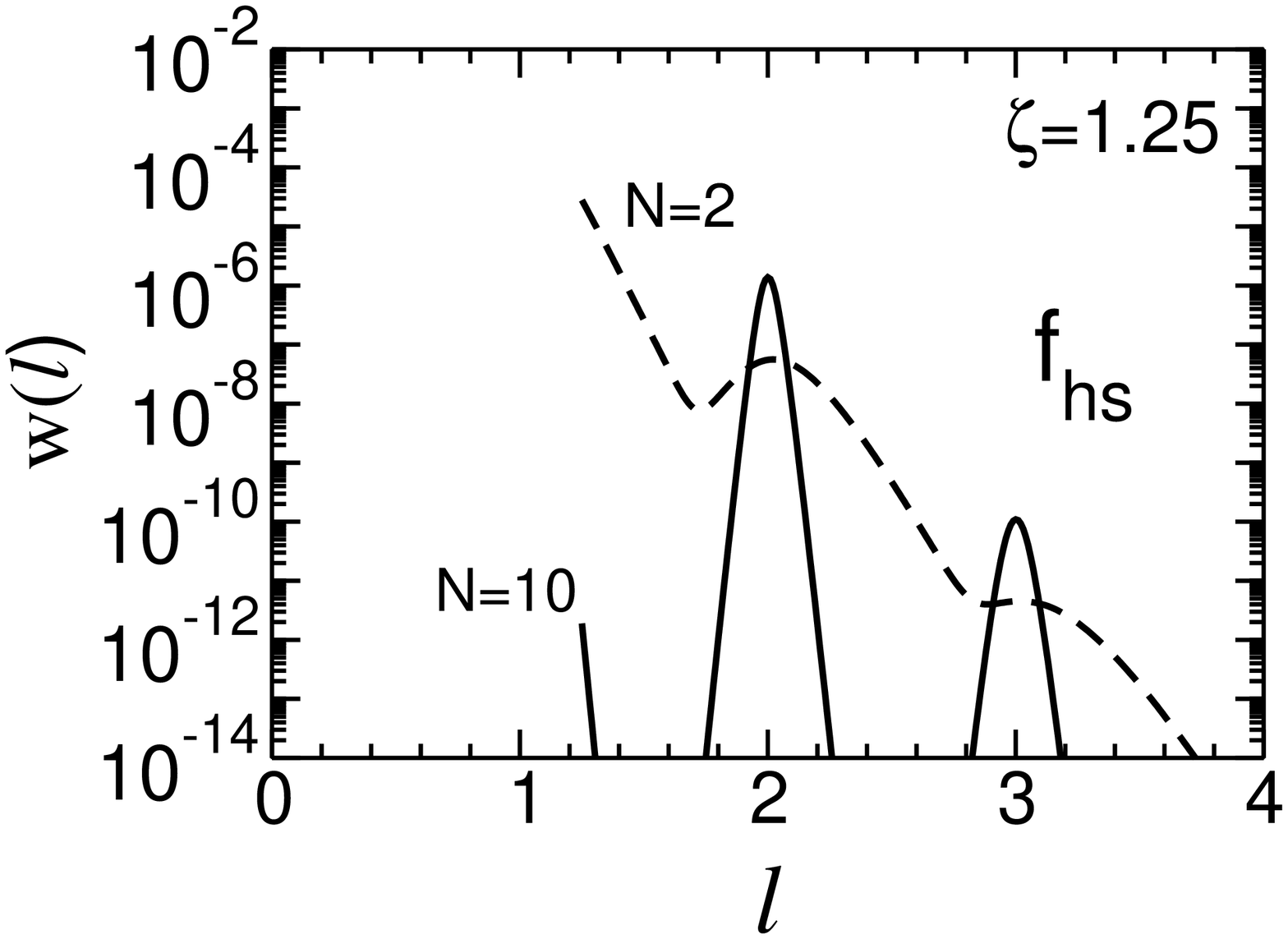}\qquad
\includegraphics[width=0.35\columnwidth]{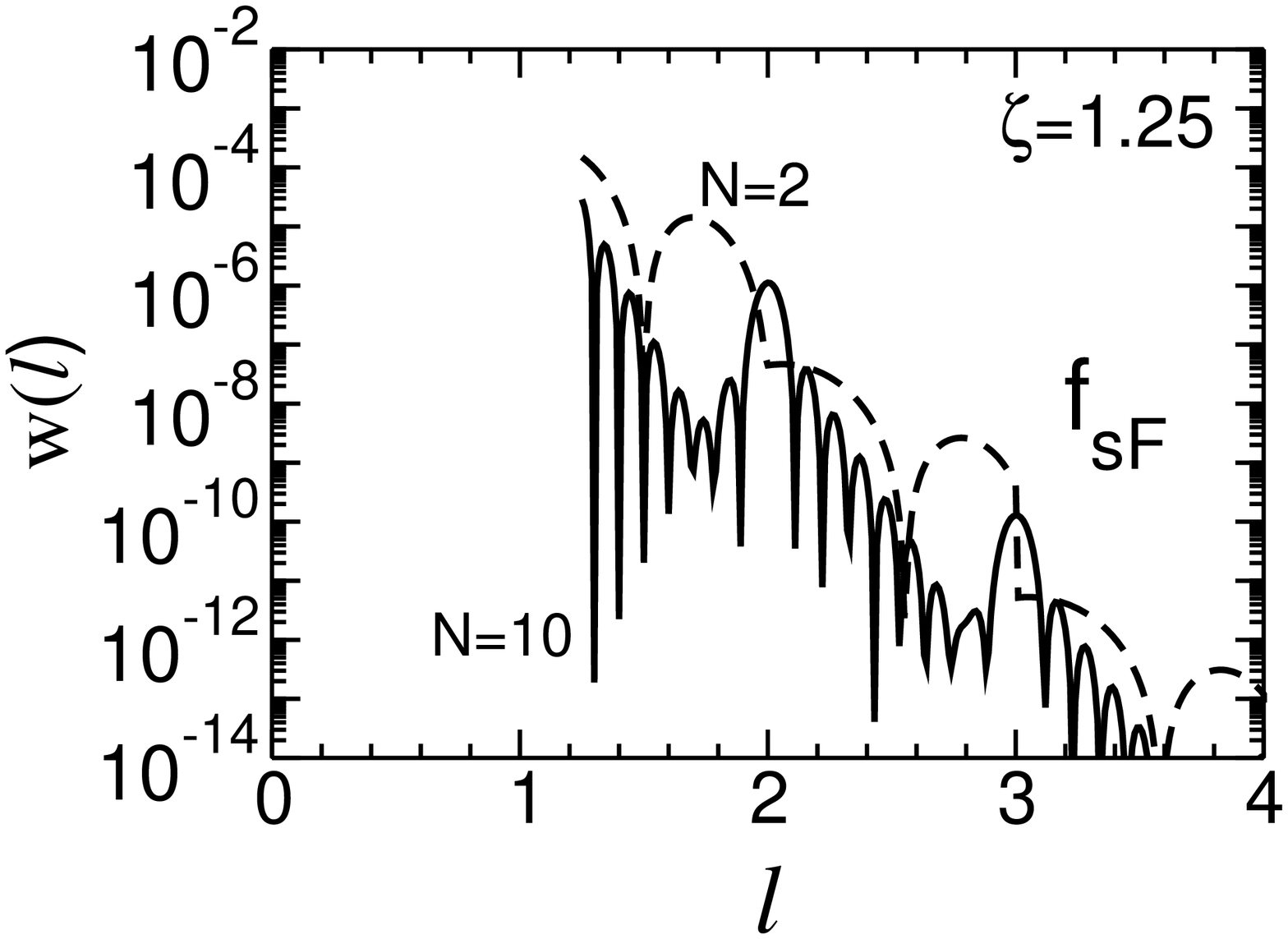}
\caption{\small{The same as in Fig.~\ref{Fig:9} but
for the sub-threshold region at $\zeta=1.25$.
 \label{Fig:10} }}
\end{figure}

In Fig.~\ref{Fig:10} we show the partial probability $w(l)$ in
the sub-threshold region,
i.e.~$\zeta=1.25$.
One can see that for the hyperbolic secant
envelope (left panel)
the difference of $w(l)$ at $l\simeq \zeta$ for $N=2$ and $N=10$ is
more than several
orders of magnitude, which will be reflected in the total probability.
In the case of the symmetrized Fermi envelope shape, one also can see
a significant enhancement of $w(l)$ for $N=2$ compared to $N=10$.
But now, the difference between FPA and IPA is larger
compared to the case of the hyperbolic secant shape.

\begin{figure}[h!]
\includegraphics[width=0.35\columnwidth]{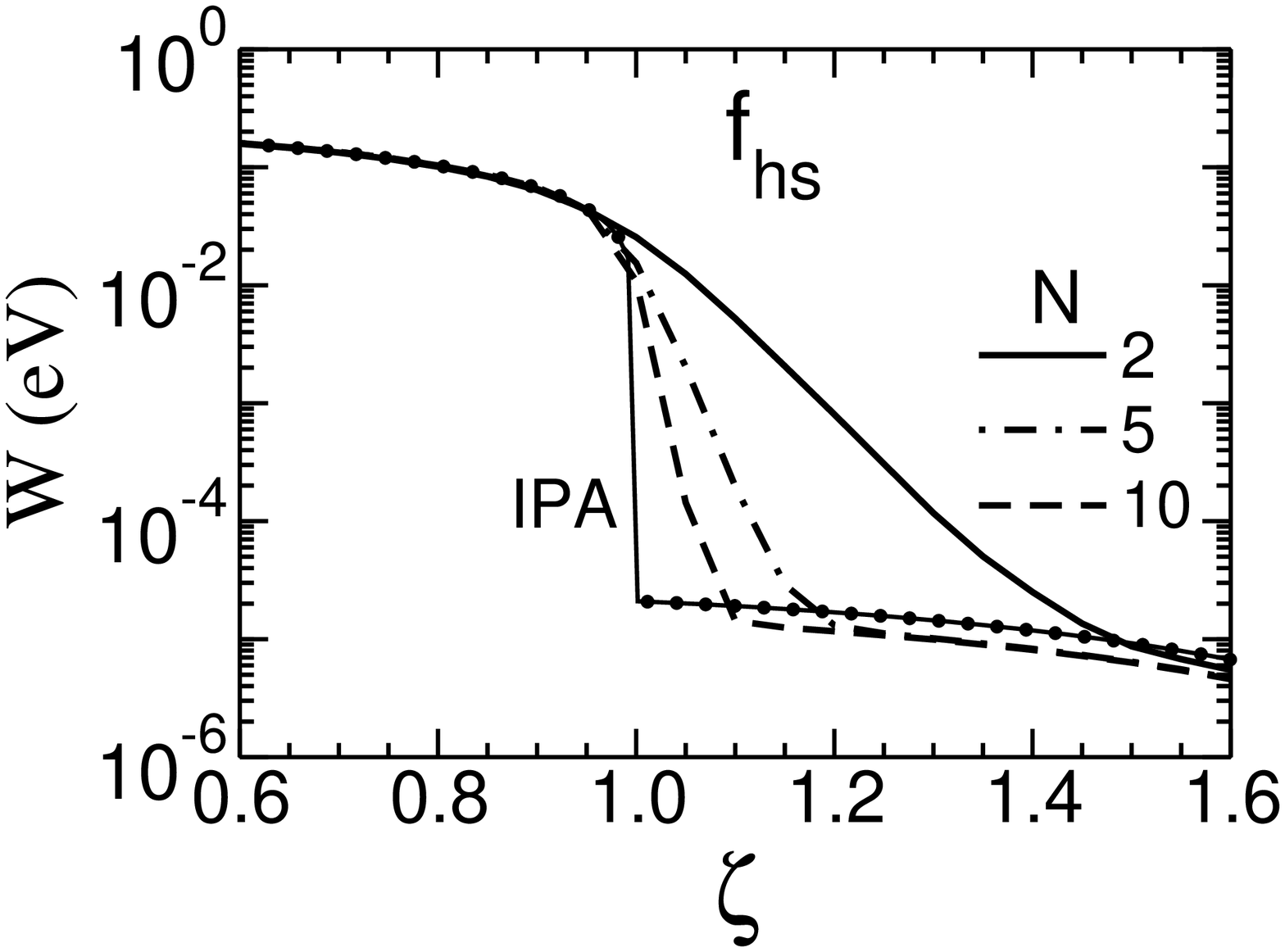}\qquad
\includegraphics[width=0.35\columnwidth]{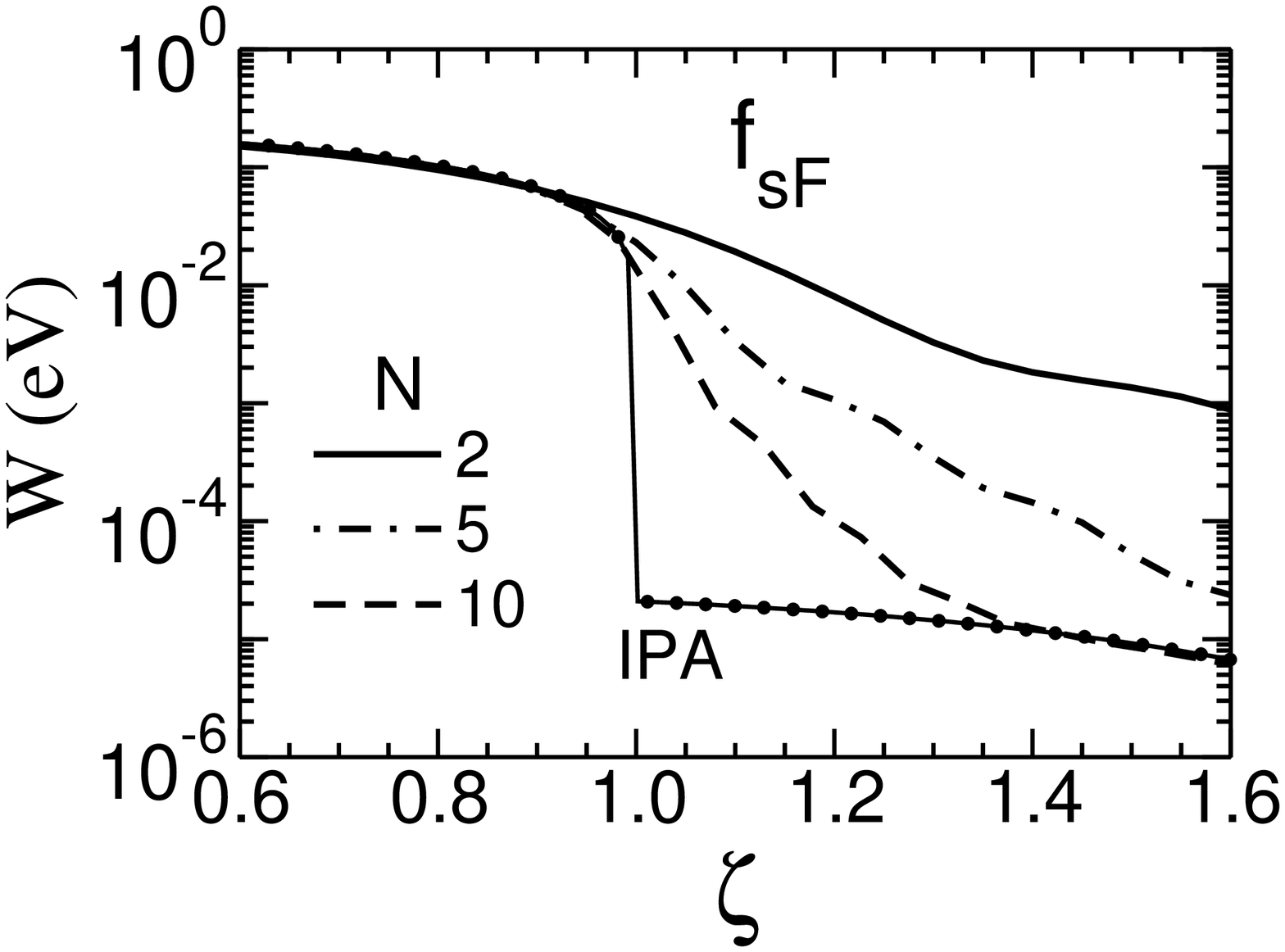}
\caption{\small{
The total probability $W$ of the
$\ee$ pair production as a function of $\zeta$
for short pulses with $\Delta=\pi N$ for $N=2$, 5 and 10
as indicated in the legend.
The thin solid curves marked by dots depict the IPA
result.
Left and right panels
correspond to the hyperbolic secant and symmetrized Fermi
envelope shapes, respectively.
 \label{Fig:11} }}
\end{figure}

\begin{figure}[h!]
\includegraphics[width=0.35\columnwidth]{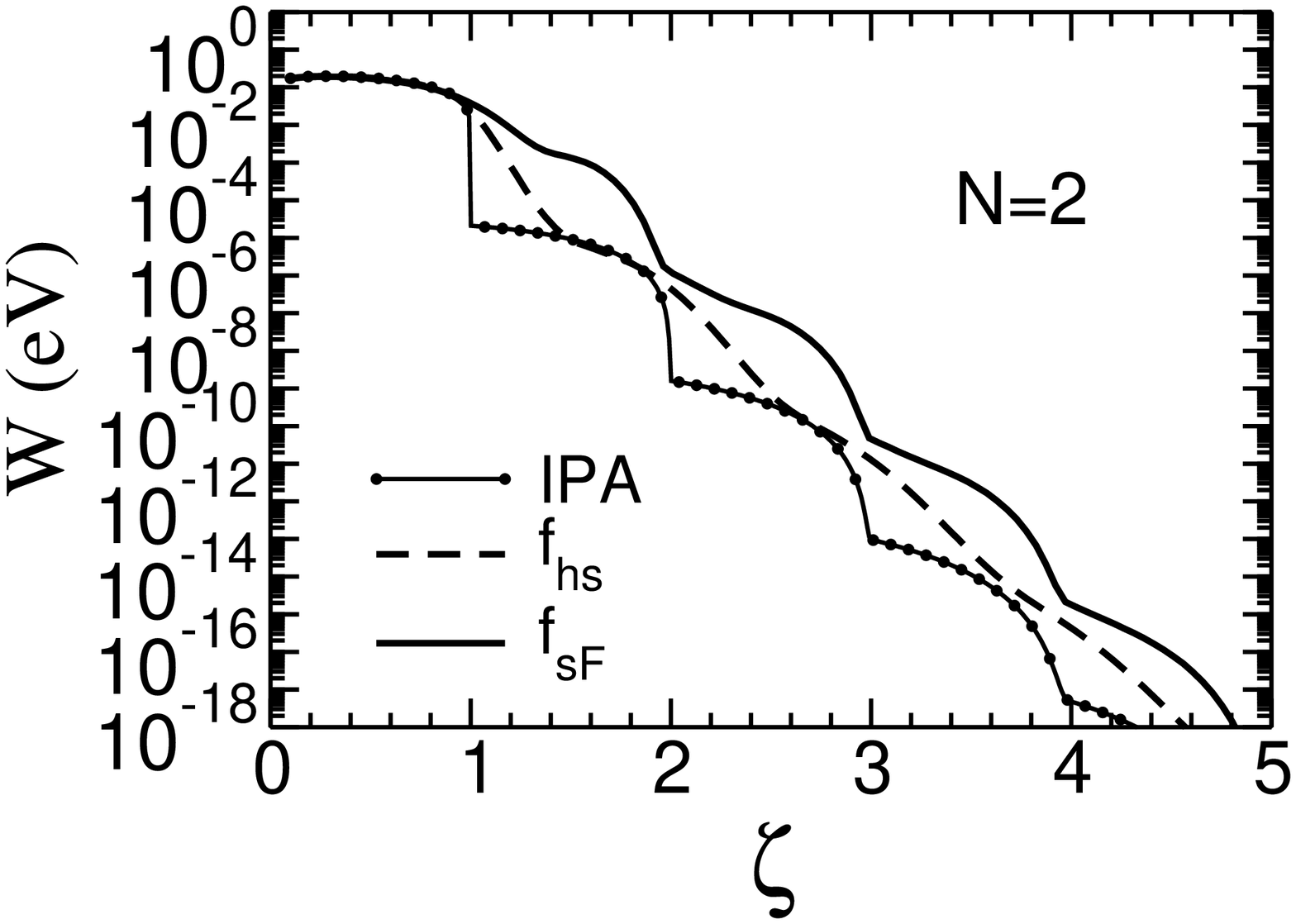}\qquad
\includegraphics[width=0.35\columnwidth]{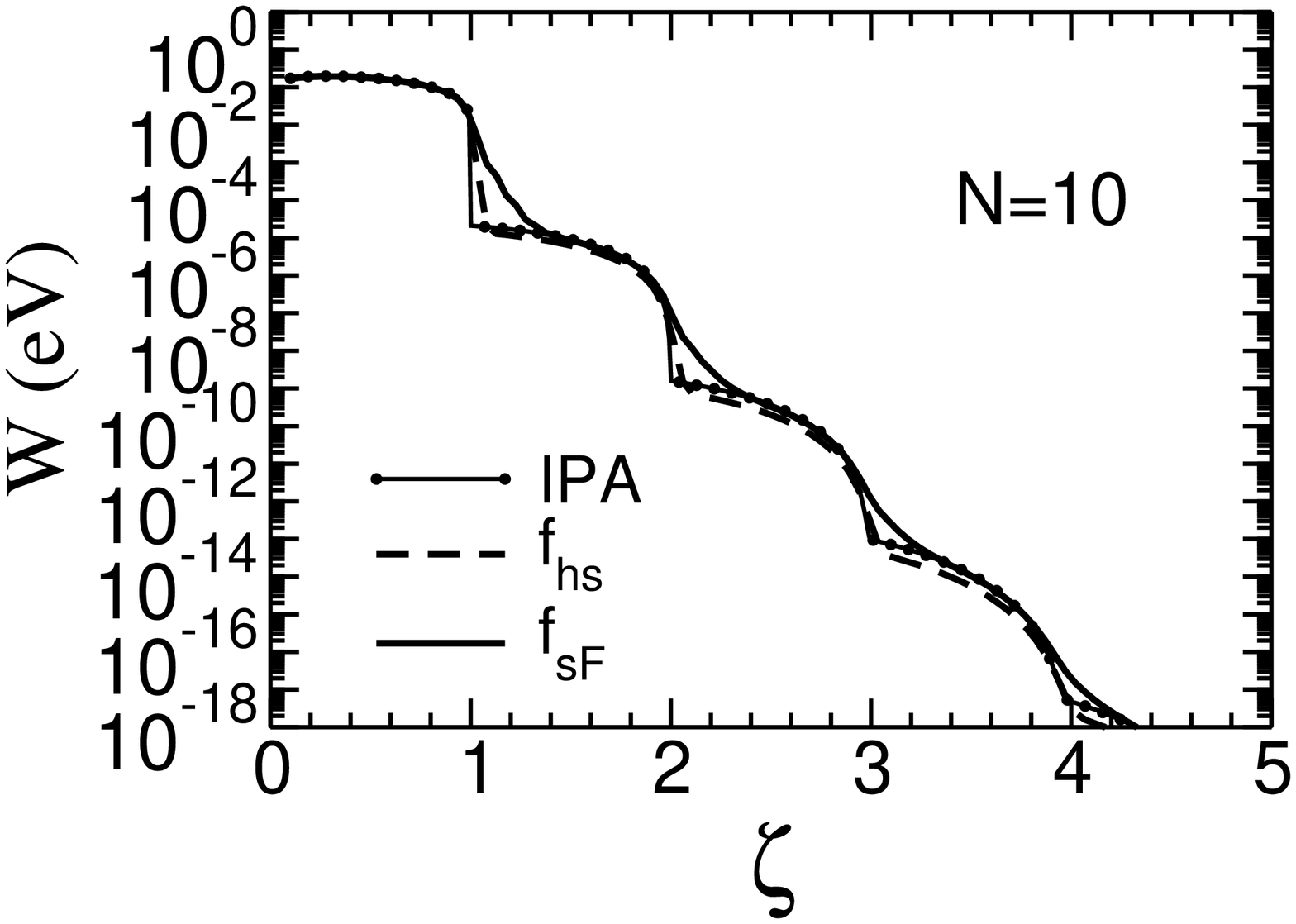}
\caption{\small{
The total probability $W$ of the
$\ee$ pair production as a function of $\zeta$
for one- and two-parameter envelope shapes
(dashed and solid curves are
for hyperbolic secant and
symmetrized Fermi shapes, respectively).
The thin solid curves marked by dots depict the
IPA result.
Left and right panels
correspond to the number of oscillation in a pulse $N=2$ and 10,
respectively.
 \label{Fig:12} }}
\end{figure}

The total probability $W$ of $\ee$ emission
as a function of the sub-threshold parameter $\zeta$
in the vicinity $\zeta\sim 1$
is presented in Fig.~\ref{Fig:11}.
The left and right panels correspond
the hyperbolic secant and symmetrized Fermi
envelope shapes, respectively.
Calculations are performed
for  short pulses with $N=2$, 5 and 10 oscillations in the pulse
at $\xi^2=10^{-3}$.
For comparison, we present also the IPA results.
In the above-threshold region, results of
IPA and FPA are equal to each other according to
Eqs.~(\ref{III28}) and  (\ref{III29}). However,
in the sub-threshold region, where $\zeta$ is close to unity,
the probability of FPA considerably
(by more than two orders of magnitude) exceeds the corresponding IPA
result. In case of the hyperbolic secant envelope function
the probability increases with decreasing  pulse duration.
The results of FPA and IPA become comparable at $N\geq 10$.
Qualitatively, this behavior is true for the case
of the symmetrized Fermi distribution.
However, in this case, the enhancement of the probability in FPA
is much greater.
This is due to the fact that the envelope of the maxima
in the partial probability
$w(l)$ (cf.~Fig.~\ref{Fig:10}) decreases with increasing $l$ in different ways
for different envelope shapes. In case of the
hyperbolic secant it decreases as $\exp(-\pi\Delta l)$,
whereas in case of symmetrized Fermi shape it decreases
as $\exp(-2\pi b l)$. For the latter one, at $b/\Delta=0.1$ the slope
is much smaller.
Such a strong gain of $\ee$ emission is expected
for other values of $\zeta$ when $\zeta$ exceeds an integer number.
This effect is illustrated in
Fig.~\ref{Fig:12}, where the total $\ee$ production probability
$W$ is presented in a wide region of $\zeta$ at $\xi=0.01$.
For convenience, we show also results for two different pulse shapes
simultaneously.
For two oscillations in a pulse (left panel $N=2$),
for the hyperbolic secant shape one can see
a regular enhancement of the probability $W$ when $\zeta$ exceeds the
corresponding integer value.
As a result, $W(\zeta)$ in FPA is a smooth function,
while a step-like dependence of the probability appears in IPA.
For the flat-top, symmetrized Fermi distribution
at $\zeta>1$, the probability is
significantly larger than for  hyperbolic secant
pulse shape and displays a step-like behavior. The latter one,
however, is related mainly to the oscillating
nature of the corresponding
Fourier transform in (\ref{B6}).

At large values of $N$ (right panel, $N=10$)
results of FPA and IPA become
close to each other, especially for the one-parameter envelope
shapes. For this case, at least for $\xi=0.01$,
$N\simeq 10$ can be considered to be near infinite,
when considering the overall $\zeta$ dependence.
For the flat-top shape with small $b/\Delta$ the probability
in FPA is higher than the result of IPA near integer values
of $\zeta$.

To summaries this part we have to note that temporal beam
shape effects for short pulses are strong and even dominant
at small field intensities in the parameter region where
the variable $z$ is small, $z\ll1$. At finite $z$, the non-linear
dynamics of $e^\pm$ in the
strong pulse becomes essential.

\subsection{Production probability at intermediate field
 intensities ($\xi^2\sim1$)}

\begin{figure}[h!]
\includegraphics[width=0.35\columnwidth]{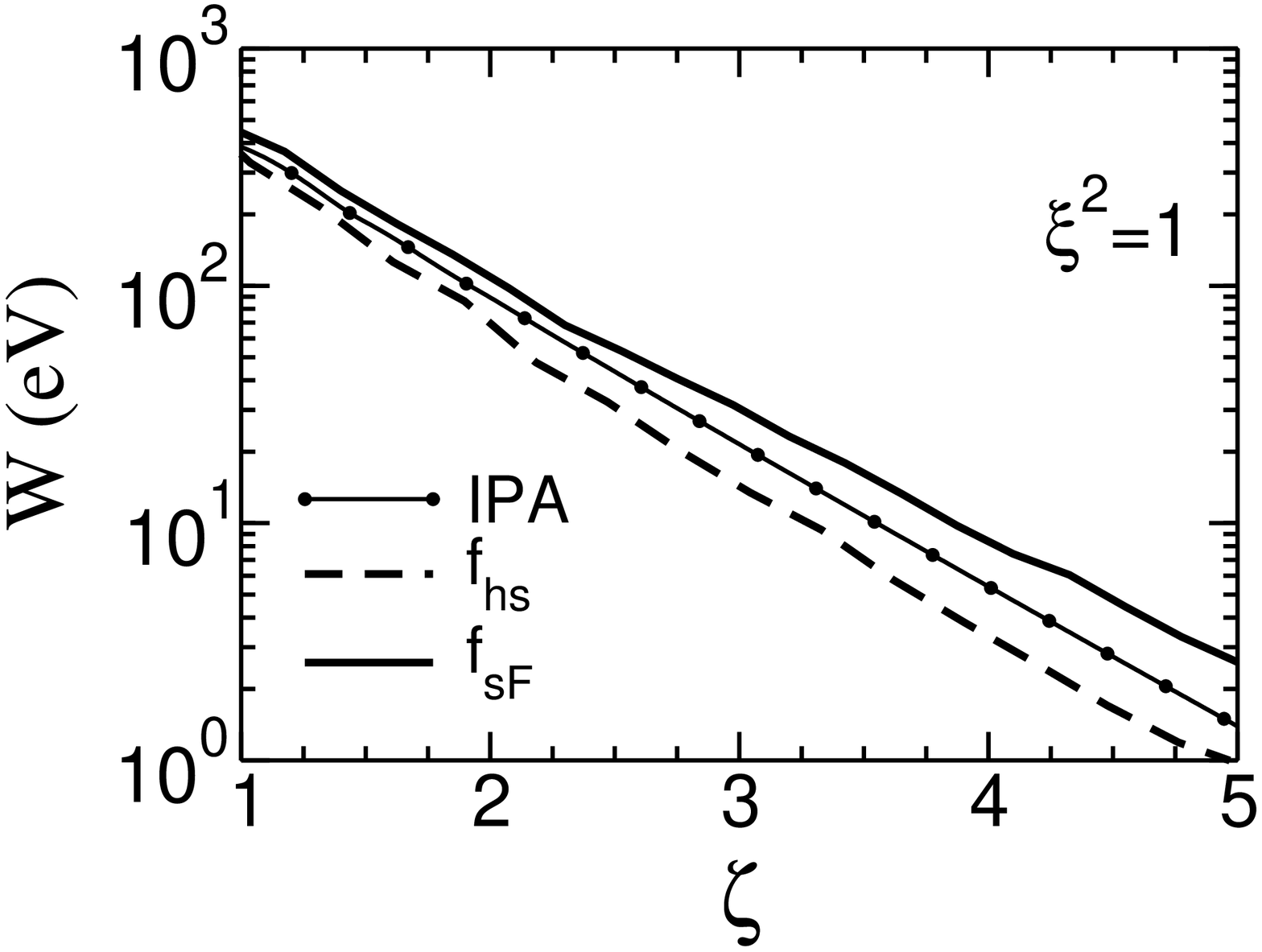}\qquad
\includegraphics[width=0.35\columnwidth]{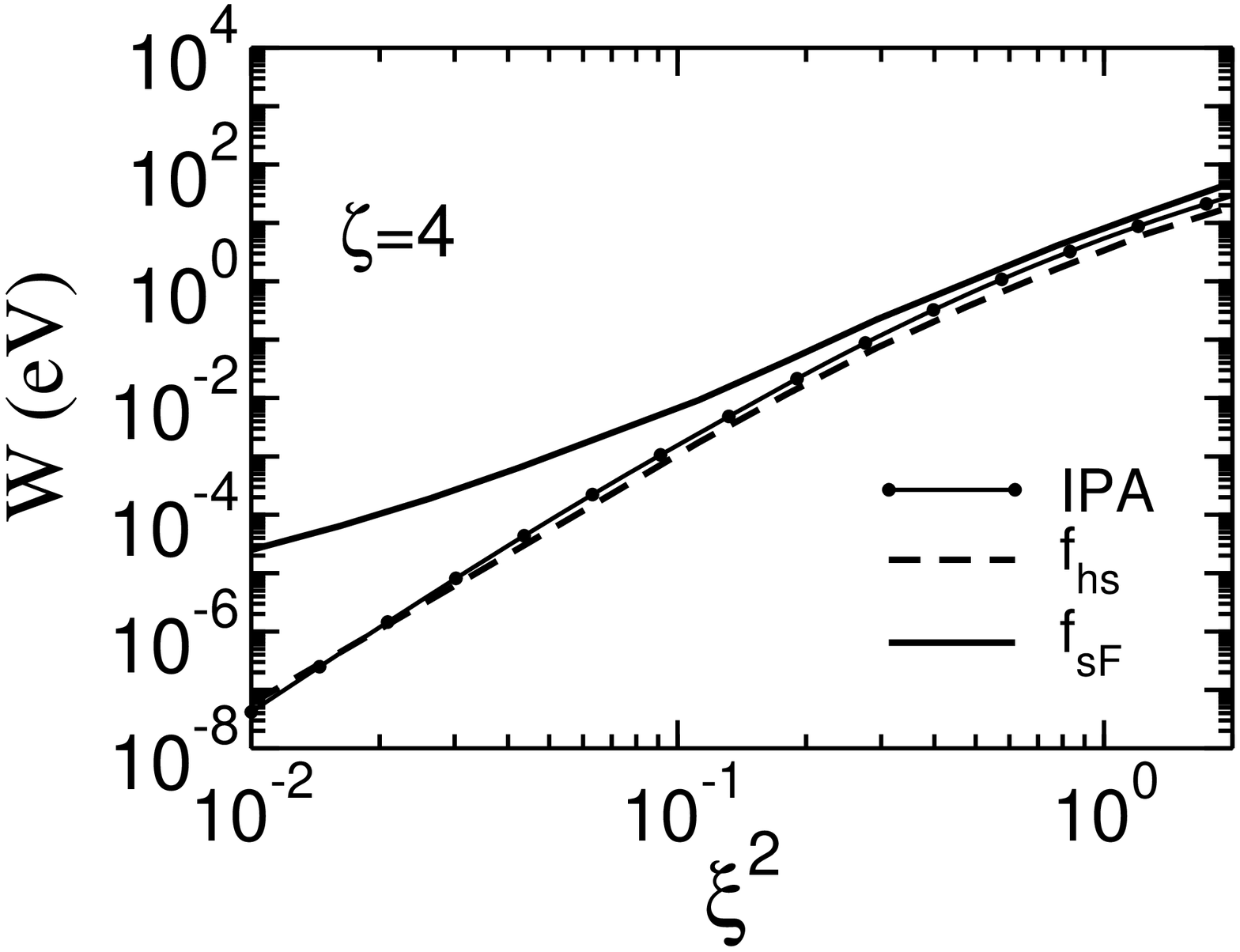}
\caption{\small{
The total probability of $\ee$-pair production
for two envelope shapes
(dashed and solid curves are for hyperbolic secant
and symmetrized Fermi shapes, respectively).
The thin solid curves marked by dots are the
result of IPA.
Left panel: The total probability as a function of $\zeta$
at $\xi^2=1$.
Right panel: The total probability
as a function of $\xi^2$ at $\zeta=4$.
 \label{Fig:13} }}
\end{figure}

At finite values of $z$, $z\gtrapprox 1$,
the probability of $\ee$ emission needs to be calculated numerically
using Eqs.~(\ref{III9}), (\ref{III24}) and (\ref{III26-0}).
In Fig.~\ref{Fig:13}, we present the total probability $W$ as
a function of $\zeta$ at fixed $\xi^2=1$ (left panel) and
as a function of $\xi^2$ at fixed $\zeta=4$ (right panel).
The calculations are performed for the hyperbolic secant and
symmetrized Fermi pulse envelope shapes,
shown by the dashed and solid curves, respectively.
The duration of the pulse is $\Delta=\pi N$ with $N=2$.
For comparison, we also present IPA results by the
thin solid curves marked by dots.
At finite $\xi^2$, the probability decreases monotonically with
increasing $\zeta$ (left panel), contrary to the step-like
decrease typical for the small $\xi^2\ll1$ (cf. Fig.~\ref{Fig:12}).
The probability for the flat-top pulse shape slightly exceeds
the probability for the hyperbolic secant and the IPA result.

Concerning the $\xi^2$ dependence (right panel),
one can see a significant enhancement of the total probability
$W$ at small values of $\xi^2$ for the flat-top pulse shape compared
to the case of hyperbolic secant and the IPA result.
The latter two results are practically identical to each other.
At $\xi^2 > 1$, the production probability does not
sensitively depend on the pulse shape, and  FPA and IPA results
are close to each other. This means that at large field
intensity the dynamical aspects of the pair production
gain a dominant role in comparison
with the pulse shape and size effects.

Finally, we note that, at finite $\xi^2$, the dependence
of the probability on the
azimuthal angle $\phi_e$ disappears and the distribution
in the $x-y$ plane becomes isotropic.
\begin{figure}[h!]
\includegraphics[width=0.35\columnwidth]{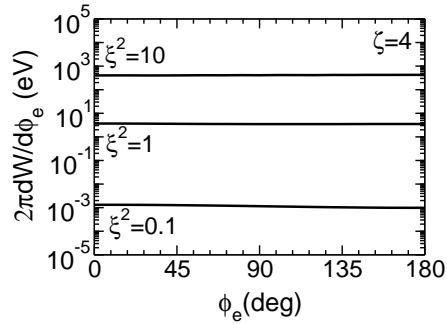}
\caption{\small{The differential probability of
$\ee$-pair production as a function of $\phi_{e}=\phi_0$
at $\zeta=4$ and $N=2$.
 \label{Fig:14}}}
\end{figure}
As an example, in Fig.~\ref{Fig:14} we present results of calculations
of the differential probability of
$\ee$-pair production as a function of $\phi_{e}=\phi_0$
at $\zeta=4$ for the hyperbolic secant pulse shape with $N=2$
at $\xi^2=0.1$, 1 and 10. The results reflect the isotropy of
the $\ee$ emission and expose the $\xi^2$ dependence.

\subsection{Production probability at large field intensity $(\xi^2\gg1)$}

At large values of $\xi^2$, $\xi^2\gg1$,
the basic functions $Y_l$ and $X_l$ in
Eq.~(\ref{III24}) can be expressed in the form of (\ref{U01}):
\begin{eqnarray}
{Y}_l=
\int\limits_{-\infty}^{\infty}dq\,F^{(1)}(q)\,G(l-q)~,\qquad
{X}_l=\int\limits_{-\infty}^{\infty}dq\,F^{(2)}(q)\,G(l-q)~,
\label{H1}
\end{eqnarray}
where $F^{(1)}(q)$ and $F^{(2)}(q)$ are Fourier transforms of
the functions $f(\phi)$ and $f^{2}(\phi)$, respectively,
and $G(l)$ may be written as
\begin{eqnarray}
G(l)=\frac{1}{2\pi}\int\limits_{-\infty}^{\infty}d\phi
\,{\rm e}^{i\left( l\phi -z\sin\phi +\xi^2\zeta u\phi
\right)}~.
\label{H2}
\end{eqnarray}
In deriving this equation we have considered the following facts:
(i) at large $\xi^2$ the probability is isotropic, therefore we put
$\phi_0=0$, (ii) the dominant contribution to the rapidly oscillating
exponent comes from the region $\phi\simeq0$, where the difference of two
large values $l\phi$ and $z\sin\phi$ is minimal, and therefore,
one can decompose the last term in the function
${\cal P}(\phi)$ in (\ref{III21}) around $\phi=0$, and
(iii) replace in exponent $f(\phi)$ by $f(0)=1$.

Equation~(\ref{H2}) represent an asymptotic form of the Bessel
functions~\cite{WatsonBook} $J_{\tilde l}(z)$
with $\tilde l= l + \xi^2\zeta u$ at $\tilde l\gg 1$, $z\gg 1$,
and therefore the following identities
are valid
\begin{eqnarray}
G(l-1) - G(l+1)=2G_z'(l), \qquad
G(l-1) + G(l+1)=2\frac{\tilde l}{z} G(l)~,
\label{H3}
\end{eqnarray}
which allow to express the partial probability $w(l)$ in
(\ref{III26-0})
as a sum of the diagonal (relative to $l$) terms: $Y_l^2$,
$Y_lX_l$, $X_l^2$ and $Y^{'2}_l$. The integral over $l$
from the diagonal term can be expressed as
\begin{eqnarray}
I_{YY}=\int\limits_\zeta^{\infty}dl\,Y_l^2=
\int dq\,dq' F^{(1)}(q)\, F^{(1)}(q')
\int\limits_\zeta^{\infty}dl
G(l-q)G(l-q')~.
\label{H4}
\end{eqnarray}
Taking into account that for the rapidly oscillating $G$ functions
$G(l-q)G(l-q')\simeq \delta(q-q')G^2(l-q)$ and
$\langle q \rangle\ll\langle l \rangle\ \sim \xi^2$ one gets
\begin{eqnarray}
I_{YY}=\frac{1}{2\pi}\int\limits_{-\infty}^{\infty}d\phi
f^{2}(\phi)\int\limits_\zeta^{\infty}dl G^2(l)
=N_{YY}\int\limits_\zeta^{\infty}dl G^2(l)~.
\label{H44}
\end{eqnarray}
Similar expressions are valid
for the other diagonal terms with own normalization factors.
For  the $X^2_l$ term it is $N_{XX}=\frac{1}{2\pi}\int\limits_{-\infty}^{\infty}d\phi
f^{4}(\phi)$, and for $Y_lX_l$,  $N_{YX}=\frac{1}{2\pi}\int\limits_{-\infty}^{\infty}d\phi
f^{3}(\phi)$. At large $\xi^2$,
the probability does not depend on the
 envelope shape, because only the central part of
the envelope is important. Therefore, for simplicity,
we choose the flat-top shape
with $N_{YY}=N_{YX}=N_{XX}=N_{0}=\Delta/\pi$ which is valid for any
smooth (at $\phi\simeq 0$) envelopes.

Making a change of the variable $l\to \tilde l= l+\xi^2\zeta u$
the variable $z$ takes the following form
\begin{eqnarray}
z^2=4\xi^2\zeta^2\left(uu_l -u^2\right)
=\frac{4\xi^2l_0^2}{1+\xi^2}\left(uu_{\tilde l} -u^2\right)
\label{H5}
\end{eqnarray}
with $l_0=\zeta(1+\xi^2)$ and $u_{\tilde l}\equiv {\tilde l}/{l_0}$,
that is exactly the same as the variable $z$ in IPA with the substitution
$l \to \tilde l$.
All these transformations
allow to express the total probability in a form similar to the
probability in IPA for large values of $\xi^2$ and a large number of
partial harmonics $n$, replacing sum of $n$ by an integral
over $n$~\cite{Ritus-79}
\begin{eqnarray}
W=\frac12 {\alpha M_e\zeta^{1/2}}
\int\limits_{l_0}^{\infty}d\tilde l
\int\limits_1^{u_{\tilde l}}\frac{du}{u^{3/2}\sqrt{u-1}}
\left\{
J^2_{\tilde l}(z) +\xi^2(2u-1)\left[
(\frac{{\tilde l}^2}{z^2} -1  ) {J}^2_{\tilde l}(z) + { J'}^2_{\tilde l}(z)
\right]\right\}~.
\label{H6}
\end{eqnarray}

Utilizing Watson's representation~\cite{WatsonBook}
for the Bessel functions at $\tilde l,\,z\gg1$ and
$\tilde l>z$,
$J_{\tilde l}(z)=({{2\pi\tilde l\tanh\alpha}})^{-1/2}
\exp[-\tilde l(\alpha -\tanh\alpha)]$ with
$\cosh\alpha={\tilde l}/{z}$,
and employing a saddle point approximation in the integration in
(\ref{H6}) we find the total
probability of $\ee$ production as (for details see Appendix~A)
\begin{eqnarray}
W=\frac{3}{8}\sqrt{\frac32} \frac{\alpha M_e \xi}{\zeta^{1/2}}
\,d\,
\exp\left[ -\frac{4\zeta}{3\xi}(1-\frac{1}{15\xi^2}) \right],
\,\,d=1+ \frac{\xi}{6\zeta}\left(1+\frac{\xi}{8\zeta} \right)~.
\label{H7}
\end{eqnarray}
This expression coincides
with the production probability in IPA which
is the consequence of the fact that, at $\xi^2\gg1$
in a short pulse, only the central
part of the envelope at $\phi\simeq 0$ is important.
Approximating $d=1+{\cal O}(\xi/\zeta)$, the leading order
term recovers the Ritus result~\cite{Ritus-79}.

\begin{figure}[h!]
\includegraphics[width=0.35\columnwidth]{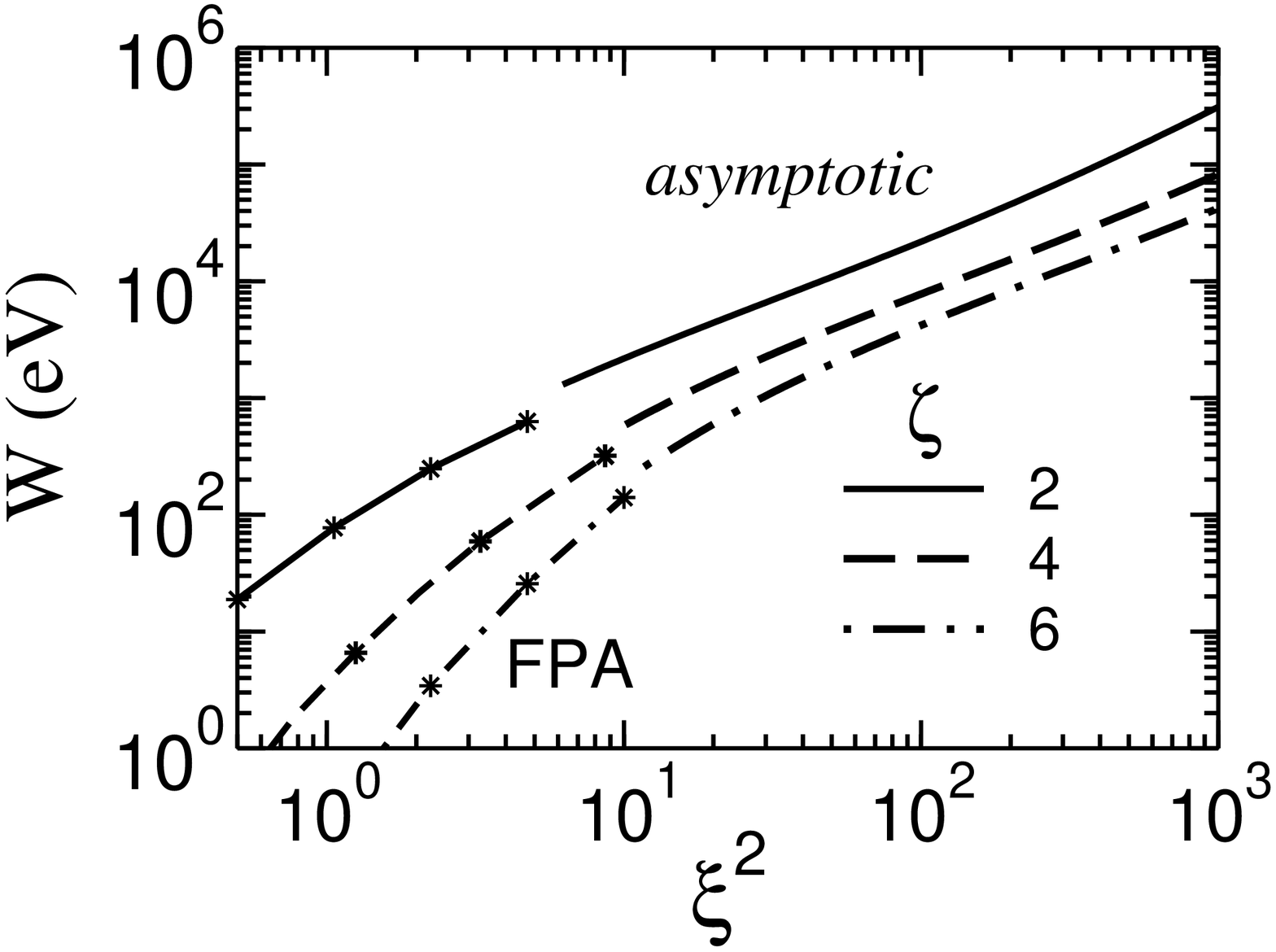}\qquad
\includegraphics[width=0.35\columnwidth]{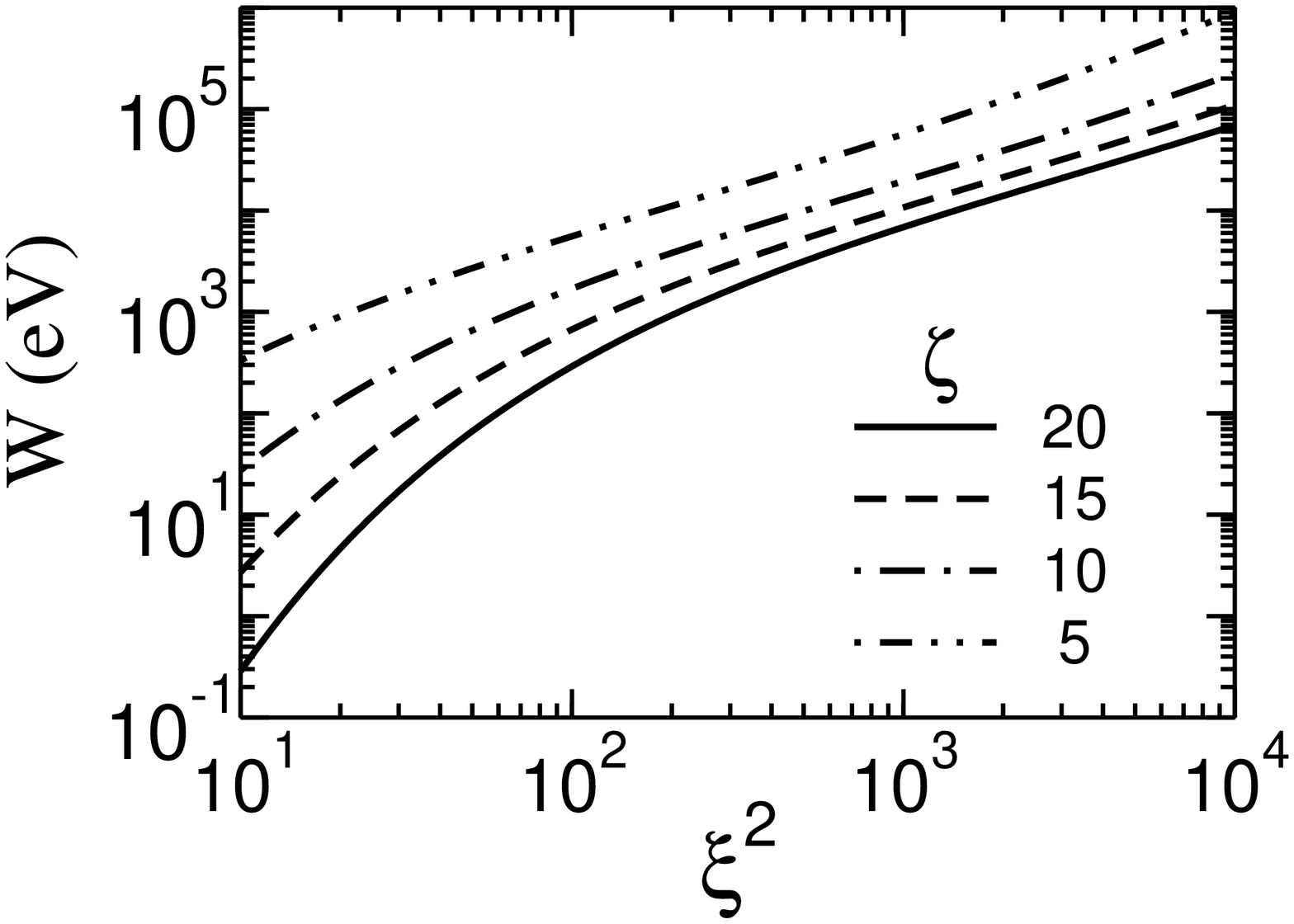}
\caption{\small{
The total probability $W$ of the
$\ee$ pair production as a function of $\xi^2$
for various values of $\zeta$.
Left panel: Results of FPA
for not too large values of $\xi^2$
(curves marked by "stars" in "FPA" sections)
and  the asymptotic probability~(\ref{H7})
for large values of $\xi^2$ (sections labeled by "asymptotic")
at $\zeta=2$, 4 and 6.
Right panel: The asymptotic probability~(\ref{H7})
for various values of
$\zeta$ as indicated in the legend.
\label{Fig:15} }}
\end{figure}

For completeness, in Fig.~\ref{Fig:15} (left panel)
we present FPA results for
not too large values of $\xi^2$
calculated for the hyperbolic secant envelope shape with $N=2$
(curves are marked by "stars")
and the asymptotic probability calculated by Eq.~(\ref{H7})
 at $\zeta=2$, 4 and 6.
The transition region between the two regimes
is in the neighborhood of $\xi^2\simeq 10$.
In the right panel, we show the production probability at
asymptotically large values of $\xi^2$
for $5 \leq \zeta\leq 20$. The exponential factor
in (\ref{H7}) is
most important at the relatively low values of $\xi^2\sim 10$
(large  ${\zeta}/{\xi}$).
At extremely large values of $\xi^2$ (small ${\zeta}/{\xi}$ ) the
per-exponential factor is dominant.


\section{summary}

In summary we have considered different aspects of $\ee$ pair production
in a strong electromagnetic field of a finite (laser) pulse,
thus generalizing the Breit-Wheeler process to
non-linear (i.e.~multi-photon) effects.
The pair production in the sub-threshold region
with $\zeta>1$ is currently a subject of great interest.
We have shown that the production probability is determined
by a non-trivial interplay of two dynamic effects. The first one is
related to the shape and duration of the pulse.
The second one is the non-linear dynamics of charged particles
in the strong
electromagnetic field itself, independently of the pulse
geometry.

These two effects play quite different roles in two limiting cases.

(i) The pulse shape effects are manifest clearly at small values
of product the $\xi \zeta$, where $\xi$ characterizes the laser
intensity and $\zeta$ refers to the threshold kinematics. The
rapid variation of the e.m. field in a very short pulse amplifies
the multi-photon events, and moreover, the probability of
multi-photon events in FPA can exceed the IPA prediction by orders
of magnitude. Thus, for example in case of an ultra-short (sub
cycle) pulse with the number of oscillations $N$ in the pulse less
than one, the production probability as a function of $\zeta$ is
almost completely determined by the square of the Fourier
transform of the pulse envelope function. High-$l$ components,
where $l$ is the Fourier conjugate to the invariant phase variable
$\phi$, lead to the enhancement of the production probability.
Among the considered envelope shapes, the flat-top shape with
small $b/\Delta$ is most promising to obtain the highest
probability. We also find that the different envelope shapes lead
to anisotropies of the electron (positron) emission which can be
studied experimentally. For short pulses with $N<10$, the effects
of the pulse shape are also important and the final yield differs
significantly from the IPA prediction. This difference depends on
the envelope shapes and the pulse duration.

(ii) Contrary to that, the non-linear multi-photon
dynamics of $e^\pm$ in the strong
electromagnetic field plays the determining role at
large field intensities, $\xi^2\gg1$.
Here, the effects of the pulse shape and duration
disappear since the dominant contribution comes from the central
part of the envelope function. As a result, the probabilities
in FPA and IPA coincide.

In the transition region of intermediate intensities
$\xi^2\sim 1$,  the probability
is determined
by the interplay of the both effects, and they must be taken
into account simultaneously by a direct numerical
evaluation of the multi-dimensional
integrals with rapidly oscillating integrands.

Finally, we emphasize  that the elaborated methods
can be applied  easily
in transport approaches aimed at studying
$\ee$ pair production
in the interaction of  electrons/positrons and/or photons
with a finite electromagnetic (laser) pulse.

\acknowledgments

The authors acknowledge fruitful discussions with
H.~Ruhr, D.~Seipt, and T.~Nousch. The support by R.~Sauerbrey
and T.~E.~Cowan is gratefully acknowledged.

\appendix

\section{Production probability at
large values of $\xi$}
The total probability $W$ in the limit of large $\xi$ and
and small $\xi/\zeta$,
was evaluated by Narozhny, Nikishov and Ritus~\cite{Ritus-79}.
for completeness and easy reference, we recall here
some details of evaluation making expansion for an
arbitrary $\xi/\zeta$.

In IPA, the
total probability is represented as an infinite sum
of partial harmonics~\cite{Ritus-79}
\begin{eqnarray}
W&=&\frac14 {\alpha M_e\zeta}\,\sum\limits_{n=n_{0}}^{\infty}
\int\limits_1^{u_n} \frac{du}{u^{3/2}\sqrt{u-1}}
 \left\{
2J^2_n(z) +\xi^2(2u-1)\left(
J^2_{n+1}(z) + J^2_{n-1}(z)-2J^2_n(z)
\right)
\right\},\nonumber\\
\label{II8-1}
\end{eqnarray}
where  $n_0\equiv n_{\rm min}=\zeta (1+\xi^2)$,
$u_n=n/n_0$, and $J_n(z)$ is the Bessel function
of the first kind  (cylindrical harmonics).
Using the identities
\begin{eqnarray}
2\,\frac{n}{z}\, J_n(z)=   J_{n-1}(z)  + J_{n+1}(z) , \,\,\,
2\,{J'}_n(z) =    J_{n-1}(z)  - J_{n+1}(z) ~,
\label{II111}
\end{eqnarray}
the total probability takes the following form
\begin{eqnarray}
W=\frac12 {\alpha M_e\zeta^{1/2}  }   \sum\limits_{n_{0}}^{\infty}
\int\limits_1^{u_n}\frac{du}{u^{3/2}\sqrt{u-1}}
\left(
J^2_n(z) +\xi^2(2u-1)\left(
(\frac{n^2}{z^2} -1  ) {J}^2_{n}(z) + { J'}^2_{n}(z)
\right)\right)~.
\label{II112}
\end{eqnarray}
At large $\xi\gg1$, $\zeta\gg1$, $n,\,z\gg1$ and $n>z$
one can replace the sum over integer $n$ by an integral over $dn$,
replacing, for convenience, integer $n$ to continues $l$ with
$l_{\rm min}\equiv l_0=\zeta(1+\xi^2)$.
Using Watson's asymptotic
expression for the Bessel functions one finds
\begin{eqnarray}
J_l\left(\frac{l}{\cosh\alpha}\right)
=\frac{1}{\sqrt{2\pi l\tanh\alpha}}
{\rm }e^{-l(\alpha   - \tanh\alpha)} +{\cal O}\left(\frac{1}{\xi}\right)
\label{II113}
\end{eqnarray}
with   $\cosh\alpha=l/z$.
If $l$ is large the first term represents a good approximation
irrespectively whether $\xi/\zeta$
is small or large~\cite{WatsonBook}.
The corresponding derivative reads
\begin{eqnarray}
J'_l(z)\simeq\sinh\alpha\,J_l(z)\,
\left(1+\frac{1}{2l\sinh^2\alpha\tanh\alpha} \right)~.
\label{II14}
\end{eqnarray}
Consider first the case of small  $\xi/\zeta\ll1$,
when the second term in (\ref{II14}) can be neglected.
Then, the total probability becomes
\begin{eqnarray}
W=\frac{e^2 M_e\zeta^{1/2} }{8\pi^2  }
 \int\limits_{l_0}^{\infty}dl\,
\int\limits_1^{u_l}\frac{du}{u^{3/2}\sqrt{u-1}}
\frac{1+2\xi^2(2u-1)\sinh^2\alpha }
{l\,\tanh\alpha}
\exp [{f(u,l)}]~,
\label{II15}
\end{eqnarray}
where $u_l=l/l_0$ and  $\hat f(u,l)=-2l(\alpha -\tanh(\alpha))$
with
\begin{eqnarray}
\tanh^2(\alpha)=\frac{1+\xi^2\left(1-   \frac{2u}{u_l}\right)^2  }{1+\xi^2}~.
\label{C2}
\end{eqnarray}
To avoid a notational confusion
with respect to the standard variable $\alpha$,
we replace below the fine structure
constant by $e^2/4\pi$.

The two-dimensional integral is evaluated using the
saddle point approximation since the function $\hat f(u,l)$ has a sharp
minimum at the point $u=\bar u$ defined by
the equation $\hat f'_u(u=\bar u)=0$.
That allows  (i) to expand it to a Taylor series
\begin{eqnarray}
f(u,l)\simeq \hat f(\bar u, l) +\frac12 \hat f_u{''}(\bar u,l)(u-\bar u)^2~,
\label{C3}
\end{eqnarray}
and (ii) to take the rest (smooth) part of the integrand in Eq.~(\ref{II15})
at the point $u=\bar u$ yielding
\begin{eqnarray}
W=\frac{e^2 M_e\zeta^{1/2} }{16\pi^2}
 \int\limits_{l_0}^{\infty}dl\,
 {\cal A}_0(\bar u, l){\rm e}^{\hat f(\bar u, l)}
 \int\limits_1^{u_l}\frac{du}{\sqrt{u-1}}
{\rm e} ^{\frac12 \hat f^{''}(\bar u,l)(u-\bar u)^2 }~,
\label{C33}
\end{eqnarray}
with
\begin{eqnarray}
{\cal A}_0(u,l) =
\frac{1+2\xi^2(2u-1)\sinh^2\alpha}
{u^{3/2}l\tanh\alpha}~.
\label{C34}
\end{eqnarray}
The explicit  expression
\begin{eqnarray}
\hat f'_u(u,l)= \frac{4l_0\sinh^2\alpha}{\tanh\alpha}
\frac{\xi^2}{1+\xi^2}
\left(1-\frac{2u}{u_l} \right)
\label{C4}
\end{eqnarray}
leads to the solution
\begin{eqnarray}
\bar u= \frac{u_l}{2}=\frac{l}{2l_0}~,
\label{C5}
\end{eqnarray}
which results in the following equalities
\begin{eqnarray}
&&\tanh\bar\alpha\equiv \tanh\alpha(\bar u)=\frac{2}{\sqrt{1+\xi^2}} , \,\,\sinh\bar\alpha=\frac{1}{\xi},
\,\,\,\hat f{''}_u(\bar u,l)=-\frac{8l_0^2}{l\sqrt{1+\xi^2}}\nonumber\\
&&{\cal A}_0=\frac{1+2(2\bar u-1)}{{\bar u}^{3/2} l}\sqrt{1+\xi^2},
\,\,\,\hat f(\bar u, l)=-2l(\bar \alpha -\tanh\bar\alpha)~.
\label{C6}
\end{eqnarray}
Using the substitutions $u=t+1$,  $a=2(\bar \alpha -\tanh\bar\alpha)$, and
 $A=-\frac12\hat f^{''}(\bar u,l)$ one can rewrite Eq.~(\ref{C33}) as
 \begin{eqnarray}
W= \frac{e^2 M_e\zeta^{1/2} }{16\pi^2 }
 \int\limits_{l_0}^{\infty}dl\,
 {\cal A}_0(\bar u, l){\rm e}^{-al -  A(1-\bar u)^2 }
 \int\limits_{0}^{\infty}dt\,  t^{\nu-1}{\rm e}^{-\beta t^2 -\gamma t} ~,
\label{C66}
\end{eqnarray}
with $\nu=1/2$, $\beta=A$, and  $\gamma=2A(1-\bar u)$.
The integral over $dt$ is expressed via the parabolic cylinder function $D_{-\nu}$
\begin{eqnarray}
\int\limits_{0}^{\infty}dt\,  t^{\nu-1}{\rm e}^{-\beta t^2 -\gamma t}
=\left( \frac{1}{2\beta}\right)^{\nu/2}\,\Gamma(\nu)\,
{\rm  exp} [\frac{\gamma^2}{8\beta}]\,
\,D_{-\nu}\left(\frac{\gamma}{\sqrt{2\beta}} \right)~,
\label{C9}
\end{eqnarray}
which results in
\begin{eqnarray}
W= \frac{e^2 M_e\zeta^{1/2} }{16\pi^{3/2} }
 \int\limits_{l_0}^{\infty}dl\,
 \left(\frac{1}{2A} \right)^{\frac14}
 {\cal A}_0(\bar u, l){\rm e}^{-al -  \frac{A}{2}(1-\bar u)^2 }
 D_{-\frac12}(y)
\label{C99}
\end{eqnarray}
with $y= \sqrt{2A}(1-\bar u)$.
The main contribution to this integral comes from
the region $\bar u\sim 1$ ($l\sim \bar l=2l_0$) and, therefore,
one can use
the  substitution
\begin{eqnarray}
\int\limits_{l_0}^\infty dl =-\frac{2l_0}{\sqrt{2A}}
\int\limits_{\sqrt{A/2}}^{-\infty}dy\approx
\frac{2l_0}{\sqrt{2A}}
\int\limits_{-\infty}^{\infty}dy~,
\label{C99_}
\end{eqnarray}
which results in
\begin{eqnarray}
W= \frac{e^2 M_e \zeta^{1/2} }{16\pi^{3/2} }
\left( \frac{1}{2A}  \right)^{\frac14}\frac{2l_0}{\sqrt{2A}}
{\cal A}_0(\bar u, \bar l){\rm e}^{-2l_0a }
 \int\limits_{-\infty}^{\infty}dy\,
 {\rm e}^{Zy -  y^2/4 }
 D_{-\frac12}(y)
\label{C999}
\end{eqnarray}
with $Z=2l_0a/\sqrt{2A}$. Using the identity
\begin{eqnarray}
 \int\limits_{-\infty}^{\infty}dy\,
 {\rm e}^{Zy -  y^2/4 }
 D_{-\frac12}(y)=\sqrt{\frac{2\pi}{Z}}\,{\rm e}^{Z^2/2}~,
\label{C999_}
\end{eqnarray}
one can rewrite the production probability as
\begin{eqnarray}
W= \frac{e^2 M_e\zeta^{1/2} }{16\pi  }
\sqrt{\frac{2l_0}{aA}}\,
{\cal A}_0(\bar u, \bar l)\,
{\exp}[ -2l_0a +\frac{l_0^2a^2}{A}]~.
\label{C9999}
\end{eqnarray}

In order to reproduce the Ritus result~\cite{Ritus-79} in terms of
the kinematic factor $\zeta$ and the field
intensity $\xi$ one has to use the identity $l_0=\zeta(1+\xi^2)$
and to represent $a(\bar \alpha)$ as a series for small values $1/\xi$
utilizing the expansions
\begin{eqnarray}
\bar\alpha={\rm arsinh}\frac{1}{\xi}
\simeq\frac{1}{\xi} - \frac{1}{6\xi^3} + \frac{3}{40\xi^5}~,\,
\tanh\bar\alpha=\frac{1}{\sqrt{1+\xi^2}}\simeq
\frac{1}{\xi} -\frac{1}{2\xi^3} +\frac{3}{8\xi^5},\,
{\cal A}_0=\frac{3}{2\zeta\xi}~
\label{C13}
\end{eqnarray}
which leads to (\ref{H7}) with $d=1$.
Inclusion of the second term in~(\ref{II14}) modifies eventually
${\cal A}_0$
as
\begin{eqnarray}
{\cal A}_0=
\frac{3}{2\zeta\xi}
\left(1+ \frac{\xi}{6\zeta}\left(1+\frac{\xi}{8\zeta} \right)\right)
\label{C14}
\end{eqnarray}
yielding the result displayed in~(\ref{H7})
which extends the Ritus result for arbitrary values of $\xi/\zeta$.
We emphasize that, in the strong field regime, IPA is representative
since, as stressed above, pulse shape and pulse duration effects are sub leading.


\end{document}